\documentclass[doublecol]{epl2} 

\usepackage{latexsym}
\usepackage{mathtools,amsmath,amssymb}
\usepackage{relsize}
\usepackage{cancel}
\usepackage{centernot}
\usepackage{halloweenmath}
\usepackage{graphicx}
\usepackage[usenames,dvipsnames]{xcolor}
\usepackage{dcolumn}
\usepackage{bm}
\usepackage{xspace}

\usepackage{color}
\definecolor{green}{rgb}{0,0.6,0}
\usepackage{hyperref}
\hypersetup{
    colorlinks,
    citecolor=blue,
    filecolor=red,
    linkcolor=green,
    urlcolor=blue
}

\usepackage[format=plain,indention=0cm,font=small, labelfont=sc]{caption}
\usepackage[font=small, labelfont=sc]{subcaption}
\newcommand{\kb}{\mathbf{k}}
\newcommand{\xb}{\bm {x}}

\newcommand{\be}{\begin{equation}}
\newcommand{\ee}{\end{equation}} 
\newcommand{\lb}{\label}

\newcommand{\wh}{\widehat}
\newcommand{\dhat}[1]{\dot{\widehat{#1}}}

\newcommand{\calR}{{\mathcal R}}

\newcommand{\bi}{{\mbox{\boldmath ${\i}$}}}
\newcommand{\bj}{{\mbox{\boldmath ${\j}$}}}
\newcommand{\bk}{{\bf k}}
\newcommand{\hBi}{\hat{\bi}}
\newcommand{\hBj}{\hat{\bj}}
\newcommand{\hBk}{\hat{\bk}}
\newcommand{\Be}{{\bf e}}
\newcommand{\hBe}{\hat{\Be}}

\newcommand{\cConv}{\circledast}
\DeclareRobustCommand{\vConv}{\overrightarrow{\cConv}}

\newcommand{\fuzConv}{\,{\mathsmaller{\mathcloud}}\,}

\newcommand{\bff}{{\bf f}}

\newcommand{\bu}{{\bf u}}

\newcommand{\bv}{{\bf v}}
\newcommand{\bx}{{\bf x}}
\newcommand{\by}{{\bf y}}

\newcommand{\bK}{{\bf K}}

\newcommand{\bY}{{\bf Y}}
\newcommand{\bPsi}{{\bf \Phi}}
\newcommand{\bPhi}{{\bf \Psi}}
\newcommand{\cS}{{{\mathbb S}^2}}

\newcommand{\Yin}{{\mathcal Y}^{\text{in}}}
\newcommand{\Yout}{{\mathcal Y}^{\text{out}}}

\newcommand{\cR}{{\mathbb R}}

\newcommand{\grad}{{\mbox{\boldmath $\nabla$}}}
\newcommand{\bdot}{{\mbox{\boldmath $\cdot$}}}
\newcommand{\GradThree}{\grad_{\!\text{\tiny{3}}}}
\newcommand{\Divt}{\grad_{\text{\!\tiny{3}}}\bdot}

\newcommand{\btimes}{{\mbox{\boldmath $\times$}}}

\title{Commutative Coarse-graining on the Sphere}
\shorttitle{Commutative Coarse-graining on the Sphere} 

\author{C. Teeraratkul\inst{1,2} \and H. Aluie\inst{1,3,4}\thanks{E-mail: \href{mailto:hussein@rochester.edu}{hussein@rochester.edu} (corresponding author) }}

\shortauthor{C. Teeraratkul and H. Aluie}

\institute{                    
  \inst{1} Department of Mechanical Engineering, University of Rochester, Rochester, NY 14627\\
  \inst{2} Division of Cardiology, David Geffen School of Medicine, University of California, Los Angeles, CA 90095\\
  \inst{3} Department of Mathematics, University of Rochester, Rochester, NY 14627\\
  \inst{4} Laboratory for Laser Energetics, University of Rochester, Rochester, NY 14627
}
\pacs{nn.mm.xx}{First pacs description}
\pacs{nn.mm.xx}{Second pacs description}
\pacs{nn.mm.xx}{Third pacs description}

\abstract{
Filtering by spatial convolution underlies coarse-graining and Large Eddy Simulation, the analysis of multiscale dynamics, and data processing across the sciences. Yet, as commonly practiced on the sphere, it violates a basic consistency requirement: commutation with differential operators, which is necessary for deriving coarse-grained dynamics and preserving the physics across lengthscales. The root of the problem is nonlocality of filtering, which compares directions in different tangent planes. The same reason underlies the Foucault pendulum's precession, and it is why the commutation failure is intrinsically geometric rather than being due to the coordinate system or numerical grid. A recently proposed generalized spherical convolution is, for a given scalar kernel, the unique linear operator commuting with the radial projection, divergence, and curl. Any consistent linear operator, be it spectral truncation, wavelet transform, diffusion, Green's-function inversion, or a machine-learned convolution, necessarily takes this form. From a simulation of an unstable jet, we demonstrate commutation to within numerical round-off, whereas common filtering that ignores curvature produces significant spurious divergence where none exists and gradient errors comparable to the gradients themselves.
}

\begin{document}

\maketitle

\section{Introduction}
A simple everyday observation already reveals the length scales at which we experience Earth's curvature: standing on a beach at sea level, the visible horizon is only about $5\,\mathrm{km}$ away. This is a reminder that ``flat space'' intuition can already fail on $\mathcal{O}(1\!-\!10)\,\mathrm{km}$ scales. While many general circulation models can be regarded as Large Eddy Simulations (LES) in spirit, there has, until recently, been
no rational framework for deriving LES equations (in physical space) for flows on the globe. The difficulty stems from the fact that the standard convolution on the 2-dimensional (2D) spherical surface $\cS$ does not commute with spatial derivatives, a property essential for preserving the physical character of flows across scales (e.g., the divergence-free or vortical character).
The main goal of this paper is to demonstrate this framework, developed in \cite{Aluie19}. As we shall see, the problem is considerably more subtle than in Euclidean (or flat) spaces and, in places, runs counter to intuition from flat geometry. The generalized convolution of \cite{Aluie19} is 
uniquely determined by commutation with radial projection, divergence, and curl: any linear operator compatible with them necessarily reduces to it. Below, we demonstrate the commutation numerically for the first time and quantify the large errors that arise from common filtering practice. Applications span geophysics, planetary and solar physics, astrophysics, cosmology, and inertial confinement fusion, as well as inverse problems and data assimilation on the sphere.

The applied mathematics literature on spherical harmonic analysis, convolutions, wavelets, and the Helmholtz decomposition of vectors and tensors on $\cS$ is vast (see \cite{Aluie19}); while our work builds on or overlaps with some of those studies, a key difference is our interest in ensuring that filtering commutes with differential operators on $\cS$, an issue that, to our knowledge, had not been addressed prior to \cite{Aluie19}. Indeed, many of these methods, including spectral truncation, wavelet transforms, diffusion-based smoothing, Green's-function inversions such as the inverse Laplacian, and the rotation-equivariant layers of neural networks with learned kernels (e.g. \cite{bonev2023spherical})
are themselves spherical convolutions with particular kernels, so the question of how they extend consistently to vectors and higher rank tensors is answered, uniquely, by the construction of \cite{Aluie19} as explained below.

As a matter of terminology, we distinguish between Cartesian coordinates and Euclidean space. One may choose non-Cartesian coordinates in Euclidean space, e.g., polar coordinates in $\cR^2$. Alternatively, one may use Cartesian coordinates while being restricted to a curved surface, such as that of a sphere embedded in flat $\cR^3$.

One might contend that, having modeled flows on spherical grids of varying resolution for decades, we already ``know'' how to write the governing equations at different scales. However, in traditional numerical modeling, resolution dependence enters the equations primarily through diffusive coefficients scaled with the grid spacing, whether empirically or
via closure assumptions. What these equations lack is a parameter
$\ell$ associated with an \emph{operator} that specifies which scales of the exact solution the simulated fields represent.
Running two simulations at different resolutions does not, by itself, provide a one-to-one correspondence between the scale content of the two solutions. 
Some form of filtering or projection must be performed on both the high-resolution solution and the governing equations. One may suggest projecting onto spherical harmonics; this suffers from two problems: (i) spherical harmonics are not a basis for regional models or for the global ocean with land boundaries, and (ii) even for the global atmosphere, which variables of the solution should we project? The most obvious choice may be the meridional and zonal (northward and eastward) velocity components but, as we explain below, this is erroneous. A general method was provided in \cite{Aluie19} that identifies the precise manner in which one must project the solution to ensure that it satisfies the coarse-grained equations on the sphere with spectral truncation being a special case. 

While the problem of commutation with derivatives due to non-uniform numerical grid discretization or the presence of boundaries has received attention in LES (e.g. \cite{ghosal1995basic}), the problem due to curvature does not seem to have been appreciated. A partial exception is the work by Jordan \cite{jordan1999large}, which makes uncontrolled approximations to reduce the problem to filtering in grid space (or token space, or logical space). In fact, this is how convolutions are performed in ``token space'' in recent machine learning (ML) applications (e.g. \cite{guillaumin2021stochastic,pathak2022fourcastnet,bretherton2022correcting,bi2023accurate}). The downside of such an approach is its dependence on a specific representation of the system (e.g., a choice of coordinate system), which does not satisfy the underlying symmetries or guarantee generalizability. For ML applications, this is akin to approximating $\sin(x)$ by polynomials: over a fixed interval $x\in[a,b]$, such an approximation converges to arbitrary precision with a sufficiently high-order polynomial (by the Weierstrass approximation theorem), yet it fails outside the ``fitting'' interval $[a,b]$. Indeed, such failure motivated \cite{bonev2023spherical} to use global spherical harmonics for ML application. Even there, however, vector fields such as winds are treated as scalar channels and are filtered componentwise (eq.~\eqref{eq:FuzzyConvVectors}).

\section{Spherical convolution}
\begin{figure*}[h!]
\centering
	\includegraphics[trim=0pt 75pt 0pt 90pt, clip,width=0.3\textwidth]{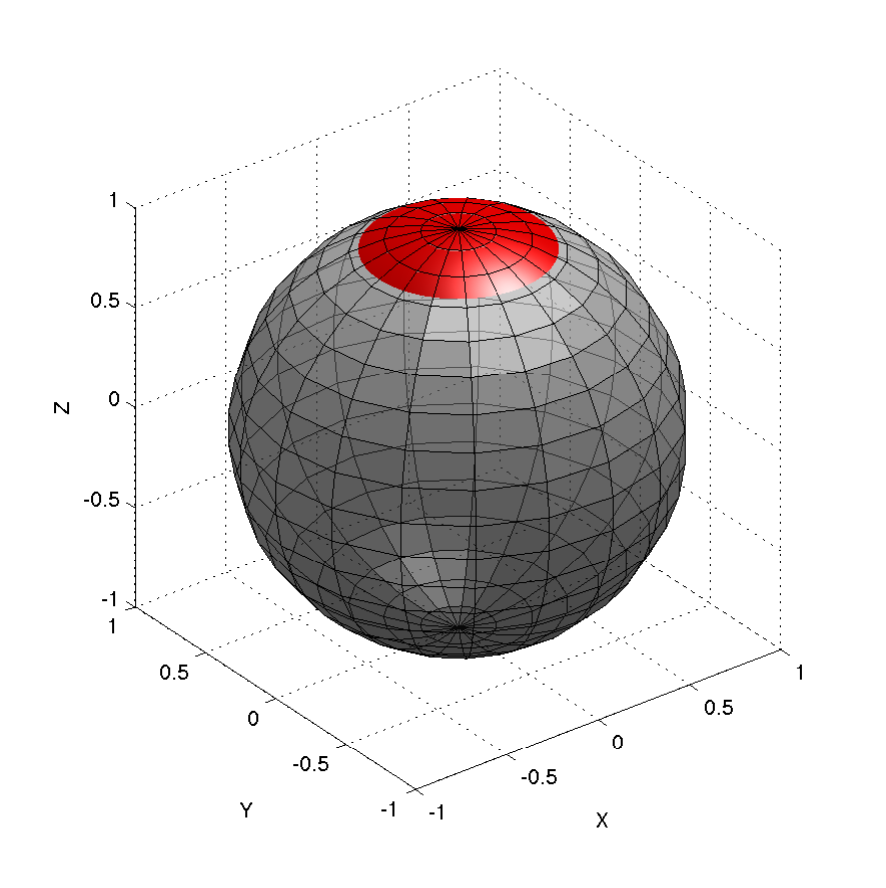}
	\includegraphics[trim=0pt 75pt 0pt 90pt, clip,width=0.3\textwidth]{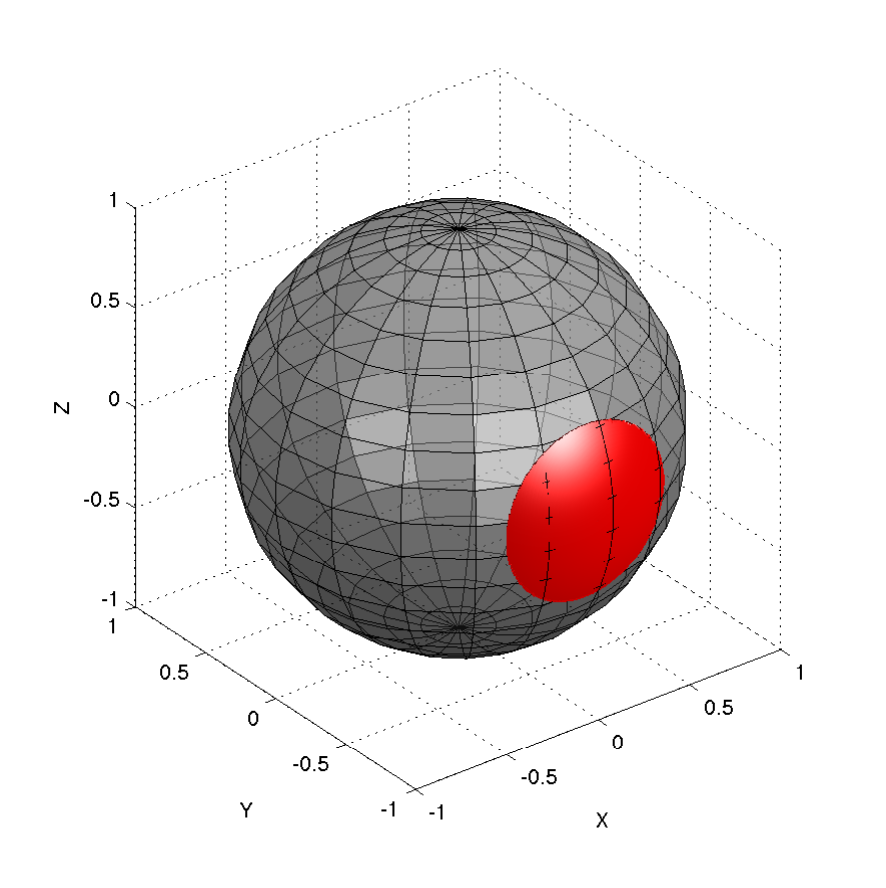}
	\includegraphics[trim=0pt 75pt 0pt 90pt, clip,width=0.3\textwidth]{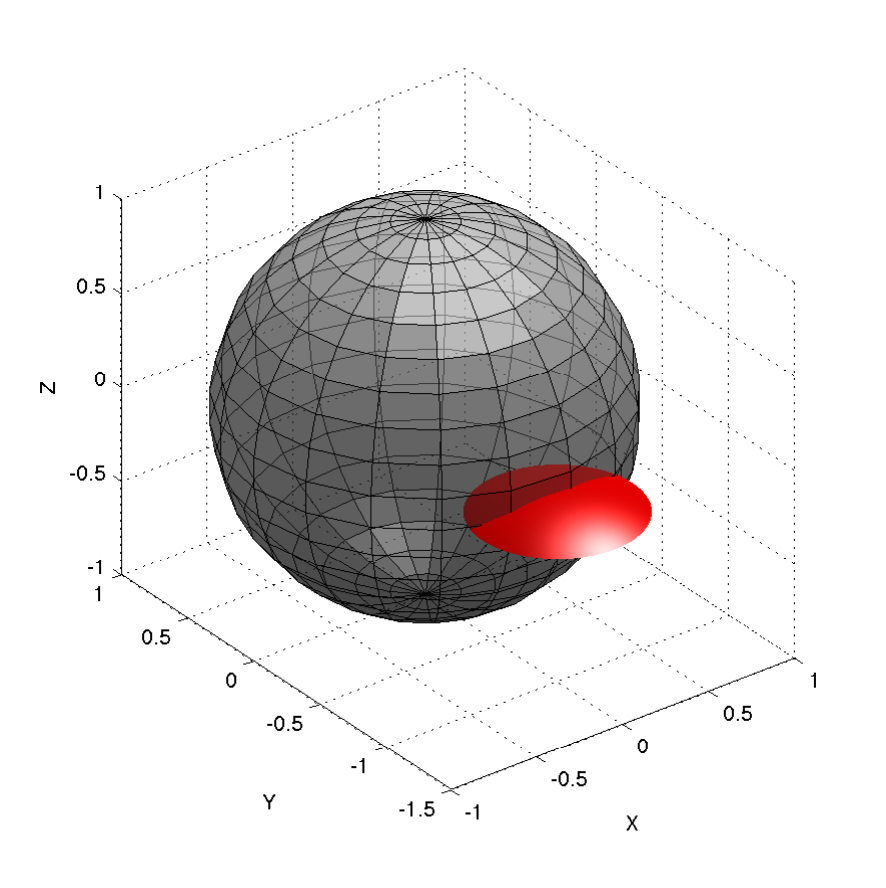}
	\caption{
Left panel shows a filtering kernel in red, e.g., a top-hat function.
Convolutions (and equivariant filtering) on the sphere are accomplished by rotating the kernel (middle panel): rotations play the role on the sphere that translations play in flat space as the underlying symmetry
group. Although the sphere is embedded in 3D Euclidean space (shown
here in units of radius $a$), translating the kernel (right panel) is
ill-defined: the translated kernel does not remain on the spherical
surface.  
}
	\label{fig:SphericalFiltering}
\end{figure*}

As Fig. \ref{fig:SphericalFiltering} shows, convolutions, and thereby filtering, on the sphere 
are accomplished by rotating the kernel, not translating it. 
For an axisymmetric kernel $G$ (about the kernel’s center, also called a ``zonal'' kernel in harmonic analysis \cite{Aluie19}, e.g., the spherical red cap in Fig.~\ref{fig:SphericalFiltering} or the Gaussian in Fig.~\ref{fig:GaussianKernel}), the spherical convolution of a scalar field $f$ can be written as
\be
G\cConv f(\bx) \coloneqq \int_{\cS} G\left(\frac{\bx\bdot\by}{a^2}\right)\, f(\by)\, dA(\by),
\lb{eq:ConvolutionScalar}\ee
where $\bx,\by\in\cS$ are position vectors, $\bx\bdot\by/a^2$ is the cosine of the geodesic angle between them, and $dA$ is the area element on $\cS$ of radius $a$.
Definition~\eqref{eq:ConvolutionScalar} and more general $SO(3)$ formulations are well known in the applied mathematics community and are routinely used in applications (see \cite{Aluie19}). However, 
fluid dynamics studies prior to \cite{Aluie19,Aluieetal18}, including those on subgrid modeling, do not seem to have appreciated that convolutions on the sphere are accomplished by rotations, utilizing instead the traditional flat convolution based on kernel translations or on averaging within grid cells. A reason may be that LES applications to geophysical flows, such as in the atmospheric boundary layer and submesoscale oceanic flows, have focused on systems of small spatial extent, $O(1-10)$~km, over which Earth's curvature is less important (e.g.,\cite{Sullivanetal14,Smithetal16}).

\section{Spherical Convolution of Vectors}
\begin{figure}[h]
	\includegraphics[trim=0pt 200pt 0pt 0pt, clip,width=0.45\textwidth]{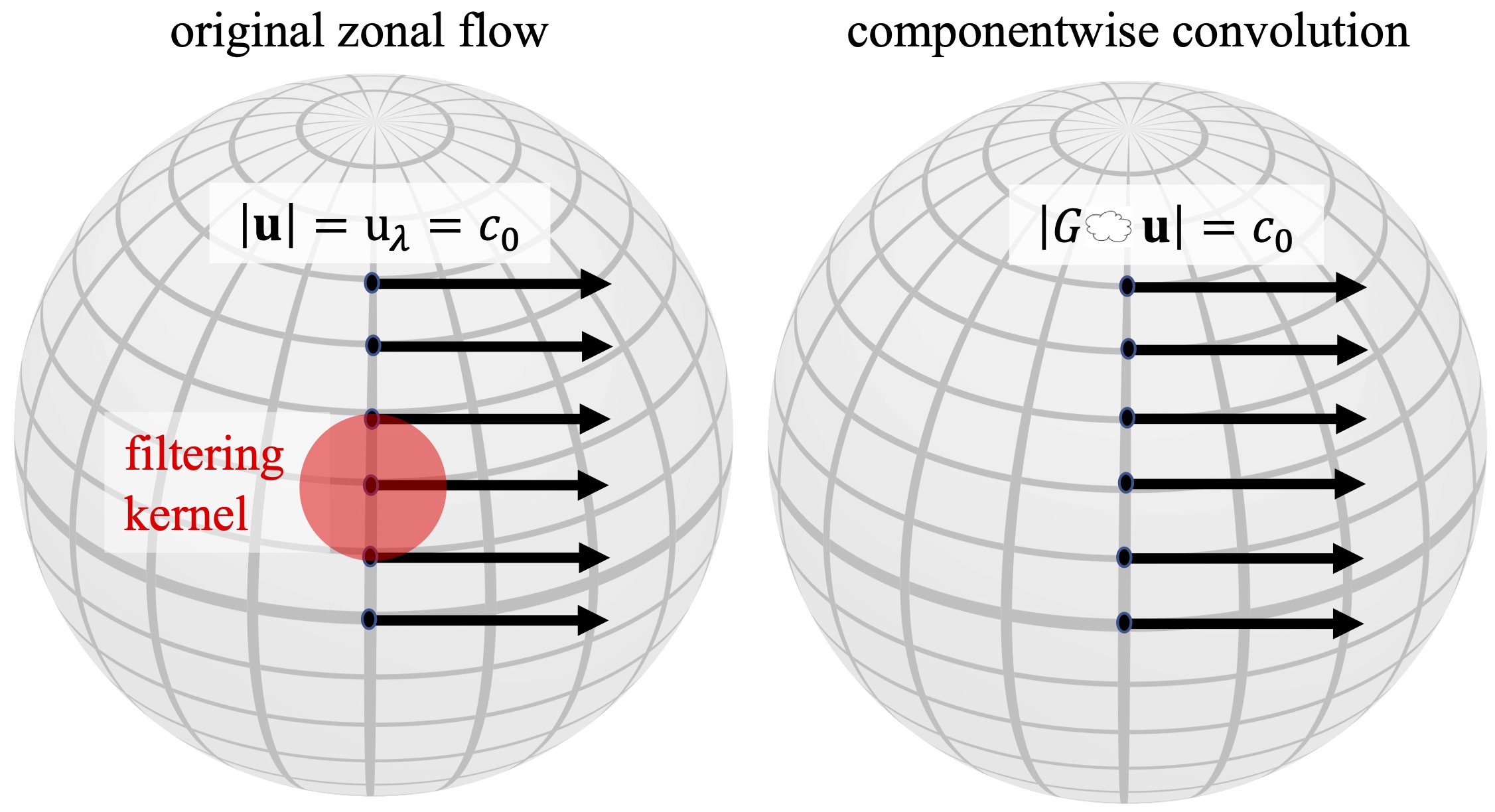}
	\caption{Componentwise convolution, i.e., locally averaging the \emph{spherical components} of a vector field (eq.~\eqref{eq:FuzzyConvVectors}), fails to decompose scales correctly on the sphere. In this example, the flow is purely zonal (eastward) with constant magnitude $|\bu| = c_0$ everywhere. On a flat manifold, such a flow would be constant, with no lengthscale except the largest possible (that of the domain). On the sphere, however, the flow has inherent gradients and, therefore, lengthscales: material elements are deformed, since the equatorial flow takes longer to complete a revolution than the flow at higher latitudes, and vorticity diverges toward the poles. Yet locally averaging the spherical components within the kernel's footprint returns the field unaltered regardless of kernel size, incorrectly implying the absence of lengthscales smaller than the domain size. 
}
	\label{fig:FuzzyConvolution_CounterZonal}
\end{figure}

In Euclidean space, the convolution of vectors and higher-rank tensors is a straightforward extension of the convolution of scalars; it is accomplished by convolving each of the tensor's Cartesian components. 
Unfortunately, many studies in geophysics, and more recently in ML-based modeling, have relied on some form of local averaging or componentwise processing of the \emph{spherical} vector components (i.e., the eastward, northward, and radial components) (e.g.\ \cite{guillaumin2021stochastic,bretherton2022correcting,pathak2022fourcastnet,bi2023accurate}).
Treating the spherical basis vectors $\hBe_r,\hBe_\lambda,\hBe_\phi$ (radial, zonal, and meridional, with $\lambda$ denoting longitude and $\phi$ latitude) as if they were spatially constant neglects both the metric terms arising from their spatial variation and, more fundamentally, the intrinsic curvature of $\cS$. We shall call a convolution acting on the spherical vector components a \emph{componentwise} convolution, 
\begin{eqnarray} 
G\fuzConv\bv \coloneqq (G\cConv v_r) \hBe_r + (G\cConv u_\lambda) \hBe_\lambda + (G\cConv u_\phi) \hBe_\phi \,,
\lb{eq:FuzzyConvVectors}\end{eqnarray}
since it acts on each spherical component separately as if it were a scalar, without attempting to respect geometry. Here, $v_r$, $u_\lambda$, $u_\phi$ are the spherical vector components of $\bv$. 
The schematic in Fig.~\ref{fig:FuzzyConvolution_CounterZonal} along with a numerical example in Fig.\,\ref{fig:SW_FilterGrad} shows how the componentwise convolution fails to respect the very notion of lengthscale, let alone commuting with differential operators. It does not even respect the underlying rotational symmetry (it is not rotation-equivariant), since $\hBe_\lambda,\hBe_\phi$ are tied to a fixed choice of pole.

It is instructive to consider the purely zonal (or eastward) flow $\bu=u_\lambda \hBe_\lambda$ in Fig.~\ref{fig:FuzzyConvolution_CounterZonal}, which has speed $u_\lambda = c_0$ (in m/s) that is spatially uniform on $\cS$. A componentwise convolution of this flow with a low-pass kernel is essentially a local averaging of $u_\lambda=c_0$, which yields the same flow unaltered. Therefore, a componentwise convolution sees the flow in Fig.~\ref{fig:FuzzyConvolution_CounterZonal} as spatially uniform, the largest scale possible, whereas the flow is multiscale  as evidenced by the deformation experienced by fluid particles and the intense vorticity near the poles (Fig.~\ref{fig:FuzzyConvolution_CounterZonal}).

\begin{figure}[h]
	\includegraphics[trim=0pt 300pt 0pt 0pt, clip, width=0.45\textwidth]{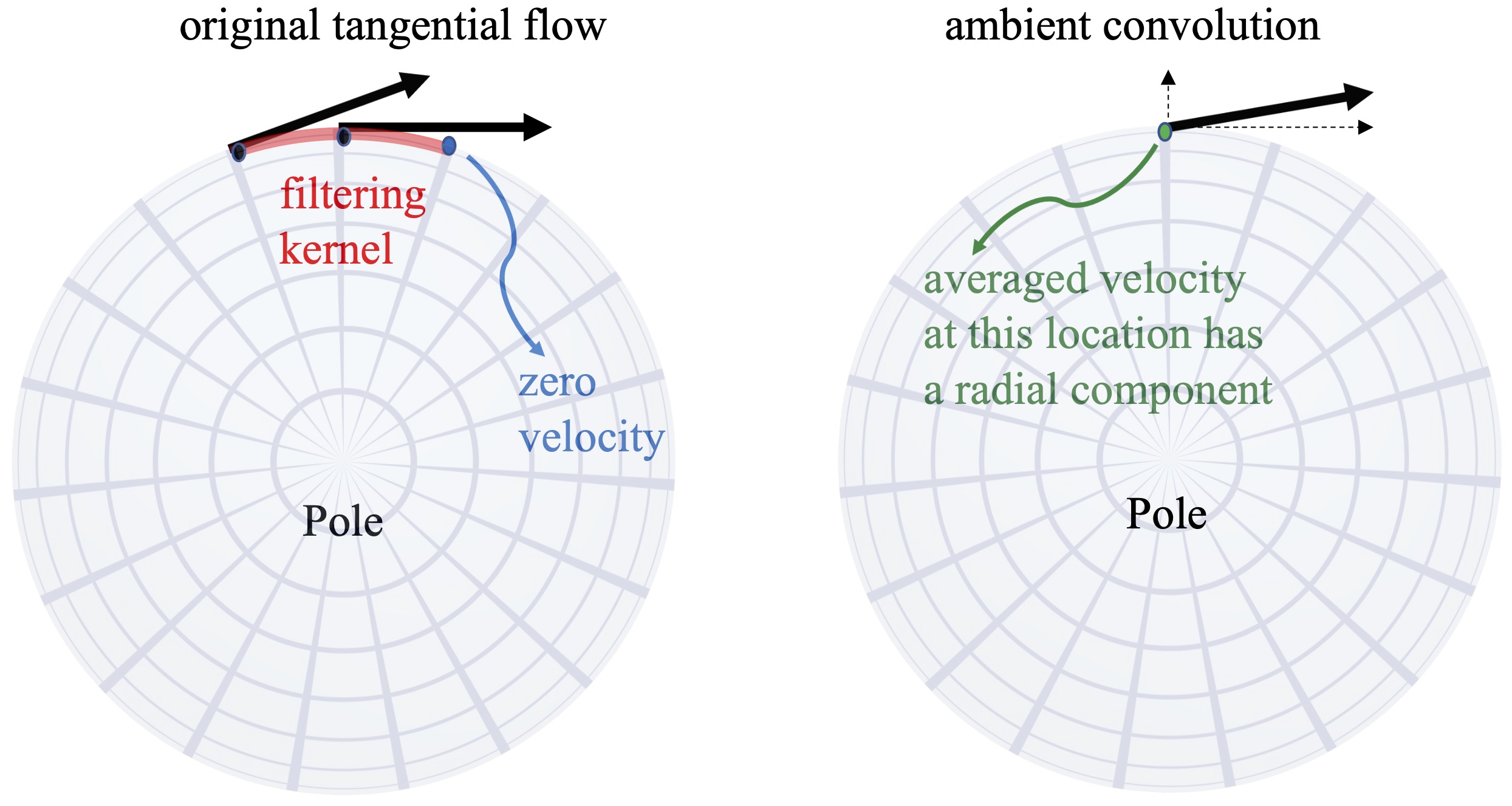}
    \caption{Consider a flow that is tangent to the spherical surface at three grid points along a great circle (the equator here), where the velocity vector has equal magnitude at two adjacent points and vanishes at the third. In this case, an ambient convolution (eq.~\eqref{eq:FacileConvVectors}) that treats the vector field in the ambient space $\cR^3$, equivalent to averaging the Cartesian components, yields a locally-averaged velocity field at the center-point that is no longer tangent to the surface but has a radial component. Therefore, the ambient convolution, despite exactly accounting for the basis vector variation (the source of metric terms), still changes the physical properties of the field. 
}
	\label{fig:SphericalFiltering_CounterPoloidal}
\end{figure}

For a laterally non-divergent (or solenoidal) flow as in Fig.~\ref{fig:FuzzyConvolution_CounterZonal}, \cite{Aluie19} showed that the \emph{ambient} convolution preserves the physical character of the flow, including its tangential and non-divergent properties. 
The ambient convolution treats the field as an object in the ambient flat space $\cR^3$, performing a local average of the vector field on $\cS$ by spherical convolution of the \emph{Cartesian} vector components: 
\begin{eqnarray} 
G\cConv\bu \coloneqq (G\cConv u_1) \hBi + (G\cConv u_2) \hBj + (G\cConv u_3) \hBk,
\lb{eq:FacileConvVectors}\end{eqnarray} 
where $u_1,u_2,u_3$ are the Cartesian components of $\bu$ and $\hBi,\hBj,\hBk$ are the fixed Cartesian unit vectors in $\cR^3$. The ambient convolution exactly accounts for the spatial variation of the spherical basis vectors, which the componentwise convolution (eq.\,\eqref{eq:FuzzyConvVectors}) ignores. Note, however, that the absence of explicit metric terms is because curvature information is contained in the
Cartesian components themselves. The ambient convolution does not account for the intrinsic curvature. This is because by going to the ambient flat space $\cR^3$,  averaging of vectors tangent to $\cS$ at different points is not, in general, tangent at the target point. This is why, for general flows that include a divergent component (e.g. Fig.~\ref{fig:SphericalFiltering_CounterPoloidal}), the ambient convolution does not commute with differential operators, e.g., 
\be\lb{eq:noncommute}
G\cConv(\grad f) \ne \grad (G\cConv f),\;\;
 G\cConv(\grad \bdot\bu) \ne \grad \bdot(G\cConv\bu).
\ee
Numerical examples in Figs.\,\ref{fig:Div3D_SW},\ref{fig:SW_FilterGrad} show how the commutation failure of eq.~\eqref{eq:noncommute} can lead to tremendous errors, exceeding $O(10^{11})$, as we discuss below. A related issue concerns purely radial and purely tangent vector fields, $f\hat{\Be}_r$ and $\bu= u_\lambda \hBe_\lambda +  u_\phi \hBe_\phi$, respectively, which do not remain radial or tangent when convolved according to (\ref{eq:FacileConvVectors}), 
\be G\cConv(f \, \hat{\Be}_r) \ne (G\cConv f) \, \hat{\Be}_r, 
\quad G\cConv(\bu \btimes \hat{\Be}_r) \ne (G\cConv\bu) \btimes \hat{\Be}_r.
\lb{eq:RadialTangentFail}\ee
Fig.~\ref{fig:SphericalFiltering_CounterPoloidal} illustrates the failure in eq.~\eqref{eq:RadialTangentFail}. 
The flow there is tangent to $\cS$ at three grid points along a great circle, with equal velocity magnitude at the middle point and one neighbor, and vanishing at the other.
In this case, an ambient convolution that treats the vector fields in the ambient space $\cR^3$ yields a locally averaged velocity at the middle point that is no longer tangent to $\cS$ but acquires a radial component. 
Even the ambient convolution, which accounts for the metric terms and respects rotational symmetry, fails to preserve the field's physical character and is ill-suited to derive tractable coarse-grained PDEs on $\cS$.

\section{Root of the Problem}
The problem lies in the nonlocality of filtering, which compares vectors in different tangent planes. There is no path-independent way to compare directions on an intrinsically curved surface. In contrast, convolutions in Euclidean space commute with derivatives because the convolution is translation-equivariant and derivatives are the generators of translations ($e^{s\frac{d}{dx}}f(x) = f(x+s)$) (e.g. \cite{goldstein2002classical}). Commutation carries through to the gradient, divergence, and curl, since these involve only the generators and a \emph{parallel} basis (the same at every point), such as $\hBi,\hBj,\hBk$, both of which pass through the convolution. Such a basis exists because in flat space a vector can be transported from one point to another unambiguously, so directions at different points can be compared. On the sphere, no such basis exists. Suppose you hold a spear pointing north at the equator and walk straight north up the $0^\circ$ meridian to the pole, where the spear points toward $180^\circ$E. Now walk sideways (without turning) down the $90^\circ$E meridian to the equator, then backward along the equator to where you began. Although all three legs are geodesics and you never turned the spear, it ends up pointing east. The same reason underlies the Foucault pendulum's precession, and it is why the commutation failure is intrinsically geometric: it reflects the curvature of the underlying manifold, not the choice of coordinates or numerical grid.
One way to see this is from an easily verifiable (perhaps not widely appreciated) fact: the Laplacian $\Delta$ on $\cS$, 
which is a (high-pass) filter, does not generally commute with spatial derivatives, e.g.\ $[\Delta,\grad]=\Delta\grad-\grad\Delta\neq 0$, a coordinate-invariant statement.
The Laplacian on $\cS$ is 
\be
\Delta f
= \frac{1}{a^2\,\cos\phi}\,\frac{\partial}{\partial\phi}
\Bigl(\cos\phi\,\frac{\partial f}{\partial\phi}\Bigr)
\;+\;
\frac{1}{a^2\,\cos^{2}\phi}\,\frac{\partial^{2} f}{\partial\lambda^{2}}\, 
\ee
for a scalar function $f(\bx)$. The tangent gradient is
\be
\grad f(\phi,\lambda)
= \hBe_{\lambda}\frac{1}{a\,\cos\phi}\,\frac{\partial f}{\partial \lambda}\,
\;+\;
\hBe_{\phi}\,\frac{1}{a}\frac{\partial f}{\partial \phi} \,.
\ee
Throughout, $\grad$, $\grad\bdot$, and $\Delta$ without subscript~`$3$' denote intrinsic operators on $\cS$.
It is easy to check that $\Delta \grad f =  \grad \Delta f -\frac{2}{a} \hat{\Be}_r\Delta f$  is not a tangent vector field in general, and has a radial component\footnote{The Laplacian, while filtering on the surface $\cS$, is acting on the vector in ambient space $\cR^3$, similar to the sketch in Fig.~\ref{fig:SphericalFiltering_CounterPoloidal}.}. 
On the other hand, $\grad \Delta f$ is necessarily tangent to $\cS$. This is why diffusion of a vector field on the sphere requires using the vector (Hodge-de Rham) Laplacian on $\cS$, $\Delta^{\text{vec}}\bu$, which maintains the field's geometric properties \cite{ebin1970groups} and commutes with the gradient, $[\Delta^{\text{vec}},\grad]=0$. On curved surfaces, the vector Laplacian is inherently distinct from the scalar Laplacian; it mixes vector components and involves additional metric-dependent terms arising from curvature. In Euclidean settings with Cartesian basis, the vector Laplacian reduces to the componentwise scalar Laplacian since there is no curvature. 
Note that the absolute (but not relative) commutation error $[\Delta,\grad]f = -\frac{2}{a}\hat{\Be}_r\Delta f$ comes mostly from the smallest scales of $f(\bx)$. 
The generalization of convolutions to vectors and tensors in \cite{Aluie19} parallels the need to define different Laplacians (filters) for tensors of different ranks.

From a symmetry perspective, the counterpart of translation on $\cS$ is rotation: an axisymmetric kernel yields a rotation-equivariant convolution, $G\cConv(f\circ \calR) = (G\cConv f)\circ \calR$ for all rotations $\calR\in SO(3)$. Equivariance guarantees commutation of $G\cConv$ (eq.\,\eqref{eq:FacileConvVectors}) with the generators of rotations (angular momentum operators $\bx\times\GradThree$, where $\GradThree$ is the gradient   in the ambient $\cR^3$) \cite{goldstein2002classical}, and hence with the scalar Laplacian, since $(\bx\times\GradThree)^2$ is $a^2\Delta$ when restricted to $\cS$. Commutation with the generators is also why the ambient convolution suffices for laterally solenoidal flows (eq.\,\eqref{eq:NOTeigenbasis_3}) \cite{Aluie19}. However, equivariance is not sufficient to guarantee commutation with the gradient or divergence on $\cS$, which, unlike their Euclidean counterparts, involve the generators and the \emph{spatially varying} basis vector $\hat{\Be}_r = \bx/a$.  The latter does not pass through the ambient convolution; instead, it shifts harmonic degree (eqs.\,\eqref{eq:EdmondsEigen},\eqref{eq:EdmondsTransf}).

To further elucidate the failure of the ambient convolution \eqref{eq:FacileConvVectors} on $\cS$, consider first a 2D flat plane embedded in $\cR^3$, in which the Fourier \emph{scalar} basis function $e^{i\kb\bdot\xb}$ is an eigenfunction of the traditional flat convolution:
\be G*e^{i\kb\bdot\xb} = \wh{G}(\kb) e^{i\kb\bdot\xb},
\ee
where $*$ denotes the traditional flat (translational) convolution and $\wh{G}(\kb)$ is the Fourier transform of the kernel.
The result carries over trivially to the Fourier \emph{vector} basis functions $e^{i\kb\bdot\xb} \hBe_m$, where $\hBe_m$ represents any of $\hBi,\hBj,\hBk$:
\be G*e^{i\kb\bdot\xb} \,\hBe_m = \wh{G}(\kb) \, e^{i\kb\bdot\xb} \,\hBe_m,
\lb{eq:FourierFlatConvEigenbasis}\ee
because Cartesian unit vectors $\hBe_m$ are spatially constant.

On the sphere $\cS$, the \emph{scalar} spherical harmonic basis function $Y_{n,j}(\lambda,\phi)$, of degree $n$ and order $j$, is similarly an eigenfunction of the spherical convolution (eq.~\eqref{eq:ConvolutionScalar}),
\be G\cConv Y_{n,j} = \dhat{G}(n) \, Y_{n,j},
\lb{eq:ScalarEigen}\ee
where $\dhat{G}(n)$ is the Legendre coefficient of degree $n$ of the axisymmetric kernel $G$. Since the Laplacian is also diagonal on $Y_{n,j}$, with eigenvalue $-n(n+1)/a^2$, the spherical convolution commutes with it, $[\Delta,\, G\cConv\,]=0$, a property that the generalized convolution below also satisfies.
However, the \emph{vector} spherical harmonics basis
$\bY_{n,j}\coloneqq Y_{n,j}\hBe_r$,
$\bPsi_{n,j}\coloneqq a\grad Y_{n,j}/\sqrt{n(n{+}1)}$,
$\bPhi_{n,j}\coloneqq a\hBe_r\btimes\grad Y_{n,j}/\sqrt{n(n{+}1)}$
are not all eigenfunctions of $G\cConv$ (eq.~\eqref{eq:FacileConvVectors}). To see this, 
define $\Yout_{n,j}$ and $\Yin_{n,j}$ as harmonic extensions of
$Y_{n,j}$ that decay over $r>a$ and are regular over $r<a$, respectively. The Edmonds basis \cite{Edmonds57}, $\bK^{(i)}_{n,j}$ for $i=1,2,3$, decomposes any vector field on $\cS$ uniquely into potential fields that are source-free outside ($\bK^{(1)}_{n,j}\propto\GradThree\Yout_{n,j}|_{r=a}$) or inside ($\bK^{(2)}_{n,j}\propto\GradThree\Yin_{n,j}|_{r=a}$) the sphere, and
non-divergent (solenoidal) fields on $\cS$ ($\bK^{(3)}_{n,j}\propto\bx\btimes\GradThree\Yin_{n,j}|_{r=a}$).
Their key property is that the \emph{Cartesian components} of $\bK^{(i)}_{n,j}$ are scalar harmonics of a \emph{single} degree, $\mathrm{deg}(i)=n{+}1,\,n{-}1,\,n$ for $i=1,2,3$ \cite{Aluie19}. By eq.~\eqref{eq:ScalarEigen}, each is then an eigenfunction of $G\cConv$,
\be
G\cConv \bK^{(i)}_{n,j} \;=\; \dhat{G}\big(\mathrm{deg}(i)\big)\,\bK^{(i)}_{n,j},
\lb{eq:EdmondsEigen}\ee
but with \emph{degree-shifted} eigenvalues for $i=1,2$. The Edmonds basis
is related to the canonical basis by a unitary map, which we call the
\emph{Edmonds transformation} \cite{Aluie19},
\begingroup
\setlength{\arraycolsep}{1pt}
\be
\hspace{-6pt}\begin{pmatrix}\bK^{(1)}_{n,j}\\[2pt] \bK^{(2)}_{n,j}\\[2pt] \bK^{(3)}_{n,j}\end{pmatrix}
\!=\!\frac{1}{\sqrt{2n{+}1}}
\!\begin{pmatrix}\sqrt{n{+}1} \!& -\sqrt{n} \!& 0\\ \sqrt{n} \!& \sqrt{n{+}1} \!& 0\\ 0 \!& 0 \!& \sqrt{2n{+}1}\end{pmatrix}
\!\begin{pmatrix}\bY_{n,j}\\[2pt] \bPsi_{n,j}\\[2pt] \bPhi_{n,j}\end{pmatrix}
\lb{eq:EdmondsTransf}\ee
\endgroup
In the Edmonds basis, $G\cConv(\cdot)$ is diagonal (eq.\,\eqref{eq:EdmondsEigen}). Conjugating $G\cConv(\cdot)$ by the Edmonds transformation \eqref{eq:EdmondsTransf} shows that
$G\cConv\bY_{n,j}$ and $G\cConv\bPsi_{n,j}$ are degree-dependent mixtures of $\bY_{n,j}$ and $\bPsi_{n,j}$ with coefficients involving
$\dhat{G}(n{\pm}1)$ (explicit expressions in \cite{Aluie19}): ambient
convolution of a purely radial field acquires a tangential component and
vice versa, which explains Fig.~\ref{fig:SphericalFiltering_CounterPoloidal} spectrally. 
Only the solenoidal basis function is unaltered, $\bPhi_{n,j}=\bK^{(3)}_{n,j}$:
\begin{eqnarray}
G\cConv \bPhi_{n,j} &=& \dhat{G}(n)\,\bPhi_{n,j},
\lb{eq:NOTeigenbasis_3}\end{eqnarray}
which explains why the ambient convolution suffices for laterally
non-divergent flows (Fig.~\ref{fig:FuzzyConvolution_CounterZonal}) but fails in general. The commutation requirement, which is uniquely satisfied by the generalized convolution \cite{Aluie19}, resolves the ambiguity in comparing directions on $\cS$.

\section{Generalized Convolution} 
With the problem formulated as in the preceding discussion, the solution comes naturally: filter each Edmonds component with a spectrally shifted kernel that undoes the degree shift. Decomposing $\bv=\bv^{(1)}+\bv^{(2)}+\bv^{(3)}$ by projection onto the $\bK^{(i)}_{n,j}$, the generalized convolution is defined as \cite{Aluie19}
\be
G\vConv\bv \coloneqq \sum_{i=1}^{3} G^{(i)}\cConv\bv^{(i)}, 
\lb{eq:GeneralConvDef}\ee
where the shifted kernels are defined via their Legendre coefficients,
\be
\hspace{-.05cm} \dhat{G}^{(1)}(n)\coloneqq\dhat{G}(n{-}1),
\;
\dhat{G}^{(2)}(n)\coloneqq\dhat{G}(n{+}1),
\;
G^{(3)}\coloneqq G,
\lb{eq:ShiftedKernels}\ee
so that every $\bK^{(i)}_{n,j}$ (and, by unitarity of
\eqref{eq:EdmondsTransf}, every $\bY_{n,j},\bPsi_{n,j},\bPhi_{n,j}$) 
is an eigenfunction of $G\vConv$ with the \emph{same} eigenvalue
$\dhat{G}(n)$.
Commutation with differential operators on $\cS$ follows
\cite{Aluie19}. 
In particular, since $G\vConv$ preserves the radial, potential, and
solenoidal characters of a vector field, a corollary in \cite{Aluie19}
shows that it can be evaluated without spherical harmonic transforms: for $\bv = \hBe_r v_r + \grad\Phi + \hBe_r\btimes\grad\Psi$ with radial
($v_r$), potential ($\Phi$), and solenoidal ($\Psi$) Helmholtz scalars,
\be
G\vConv\bv = (G\cConv v_r)\,\hBe_r + \grad(G\cConv\Phi)
+ \hBe_r\btimes\grad(G\cConv\Psi),
\lb{eq:GeneralVectorConv}\ee
which provides a practical method for computing generalized
convolutions \cite{storer2023flowsieve}. Below we evaluate both eq.~\eqref{eq:GeneralConvDef} and eq.~\eqref{eq:GeneralVectorConv} numerically.
The procedure in eq.~\eqref{eq:GeneralVectorConv} arguably obscures the nature of averaging vectors; eq.~\eqref{eq:GeneralConvDef}, formulated in the Edmonds basis, provides the insight that guided the counterexamples in Figs.~\ref{fig:FuzzyConvolution_CounterZonal},\ref{fig:SphericalFiltering_CounterPoloidal}.
We emphasize that commutativity with differential operators is a property of the convolution operation itself and does not rely on any particular kernel $G$ beyond axisymmetry: admissible kernels include low-pass, high-pass, isotropic wavelets, Green's functions (e.g., of the Laplacian, yielding its inverse), and kernels learned from data. Moreover, for a given scalar kernel, $G\vConv$ is the unique linear operator commuting with the radial projection, divergence, and curl (see Supplementary Material, SM), and the construction extends to rank-2 tensors \cite{Aluie19}, as required for coarse-graining the subgrid stress in LES.
This construction parallels the theory of Laplacians on $\cS$ but is more general: $\Delta^{\text{vec}}$-based smoothing corresponds to the heat kernel alone, whereas the generalized convolution admits arbitrary axisymmetric kernels \cite{Aluie19}. The generalized convolution stands to the ambient convolution precisely as $\Delta^{\text{vec}}$ stands to the scalar Laplacian applied to the Cartesian components, which is diagonalized by the Edmonds basis with the same degree shifts as in eq.~\eqref{eq:EdmondsEigen}.

\section{Numerical Results}
We demonstrate our results using the single-layer shallow water equations (SWE) in a co-rotating frame on a sphere. The model describes the motion of an incompressible layer of fluid with a rigid lower boundary and an upper free surface. The SWE in vector-invariant form are (e.g. \cite{krishnamurti2006introduction}):
\begin{subequations}
\begin{align}
&\hspace{-.2cm} \partial_t\bu + (\bff+\omega_r\hBe_r)\btimes\bu    
= -\grad\left(\frac{|\bu|^2}{2}+g\,h\right)  +\nu\Delta^{\text{vec}}\bu \hspace{.0cm} 
\lb{eq:SWequations_VecInvForm_1}\\
&\partial_t h + \grad\bdot(h\,\bu) = \kappa\Delta h 
\lb{eq:SWequations_2}
\end{align}
\lb{eq:SWequations}\end{subequations}
where $\bu(\lambda,\phi)$ is the velocity tangent to the surface, $\omega_r = (\grad\btimes\bu)\bdot\hBe_r$ is the scalar vorticity\footnote{The full vector $\grad\btimes\bu$ is not intrinsically defined on $\cS$ since its tangential components  require radial derivatives. However, its radial component $\omega_r$ involves only tangential derivatives. Equivalently, $\omega_r = -\grad\bdot(\hBe_r\btimes\bu)$. Only this scalar vorticity appears here.} field normal to the surface, 
$h(\lambda,\phi)$ is the fluid ``height'' (or layer thickness) field, $\bff=2\Omega\sin\phi\,\hBe_r$ is the Coriolis vector, $\Omega=7.29\times10^{-5}~$rad/s is the angular velocity, $g=9.81~$m/s$^2$ is the gravitational acceleration, $\nu=4\times10^{4}~$m$^2/$s is the viscosity, and $\kappa=\nu$ is the thickness diffusivity. $\cS$ has Earth radius $a=a_E:=6.37\times10^6$m. While the tangent velocity $\bu$ is divergent, $\grad\bdot\bu\ne0$, the 3D velocity $\bv = \bu + v_r \hBe_r$,
\be
\Divt\bv = 0,
\lb{eq:Div3D}\ee
where $\Divt$ denotes the 3D divergence in ambient space $\cR^3$.

We solve eq.~\eqref{eq:SWequations} with our pseudo-spectral code DiNuSphere, a spherical version of DiNuSUR \cite{Zhao2022} built on MPI-parallel spherical harmonic transforms via libsharp \cite{reinecke2013libsharp}; such spectral solvers are standard in atmospheric research \cite{krishnamurti2006introduction}. Time-stepping is 4th-order explicit Runge-Kutta on a Gauss-Legendre grid with
$n_\phi = 512$, $n_\lambda = 1024$, and triangular truncation
$n_{\max} = \lfloor2n_\phi/3\rfloor$ (T341).

The initial conditions for the simulation are those of a barotropically unstable zonal (eastward) flow given in \cite{Galewskyetal04}. The initial zonal flow $\bu(\bx,t=0)$, which represents a mid-latitude jet in the troposphere, is  balanced by an associated height field $h(\bx,t=0)$. A small-amplitude height field perturbation $h'(\bx,t=0)$ is introduced to trigger the instability (see \cite{Galewskyetal04}).

\begin{figure}[h]
\centering
\begin{subfigure}[t]{0.24\textwidth}
\includegraphics[width=\textwidth]{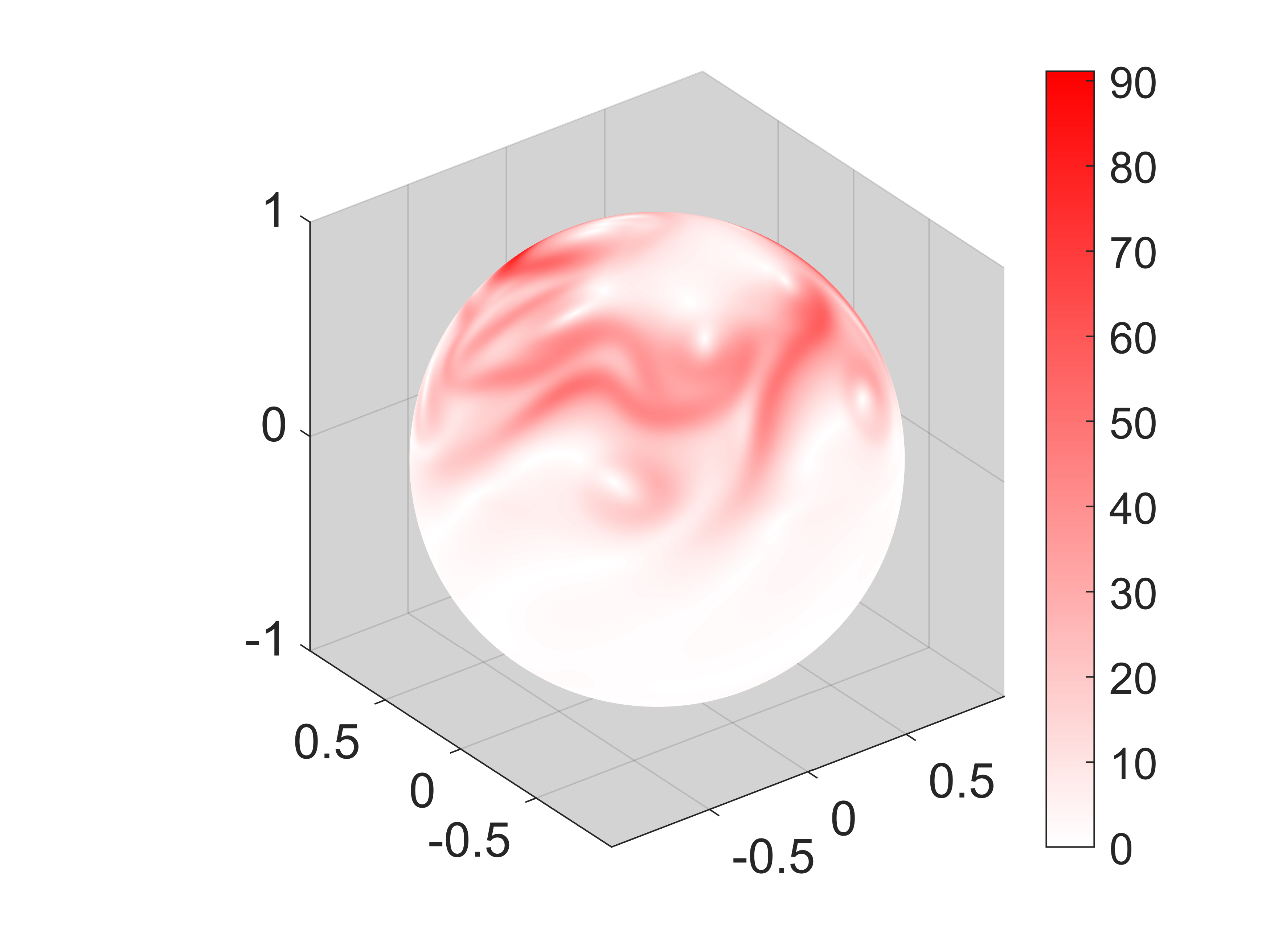}
\caption{Speed $|\bu|$ (in m/s) }
\label{fig:magnitude}
\end{subfigure}
\begin{subfigure}[t]{0.24\textwidth}
\includegraphics[width=\textwidth]{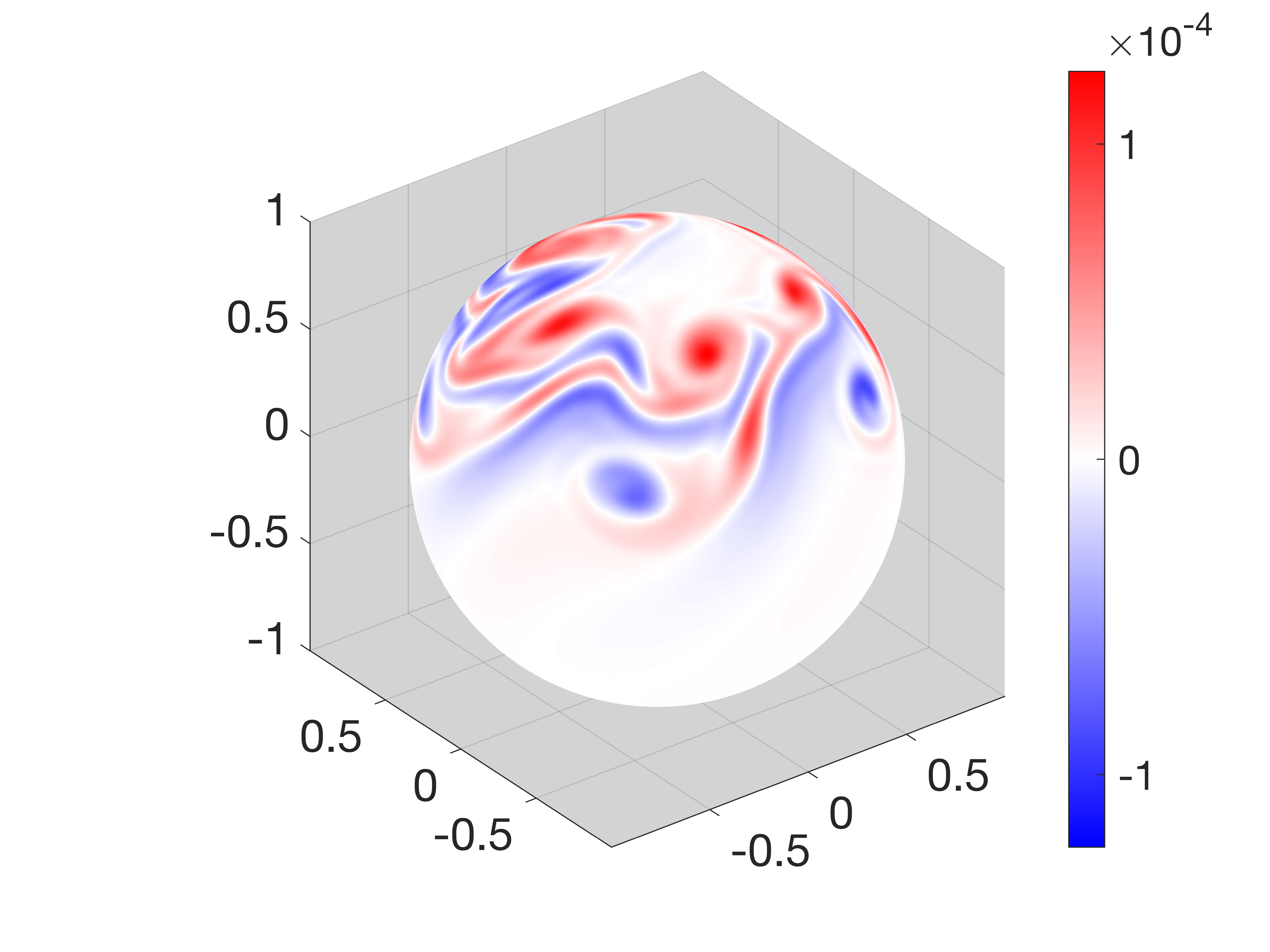}
\caption{$(\grad\btimes\bu)\bdot\hat\Be_r$ (in  s$^{-1}$) }
\label{fig:SWsimulation_vort}\end{subfigure}
\begin{subfigure}[t]{0.24\textwidth}
\includegraphics[width=\textwidth]{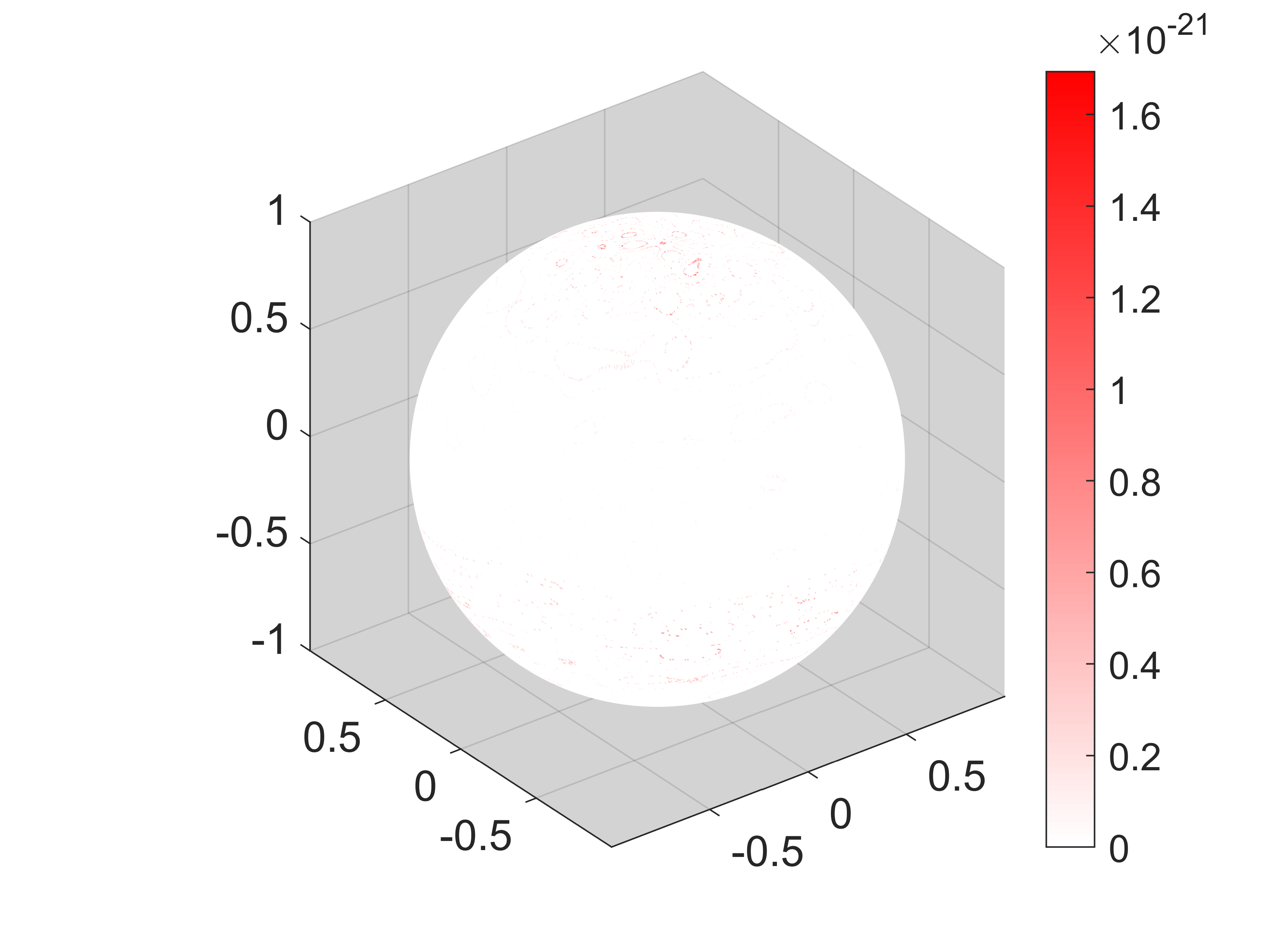}
\caption{$\Divt\bv$ (in s$^{-1}$) }
\lb{fig:SWsimulation_div}\end{subfigure}
\begin{subfigure}[t]{0.24\textwidth}
\includegraphics[width=\textwidth]{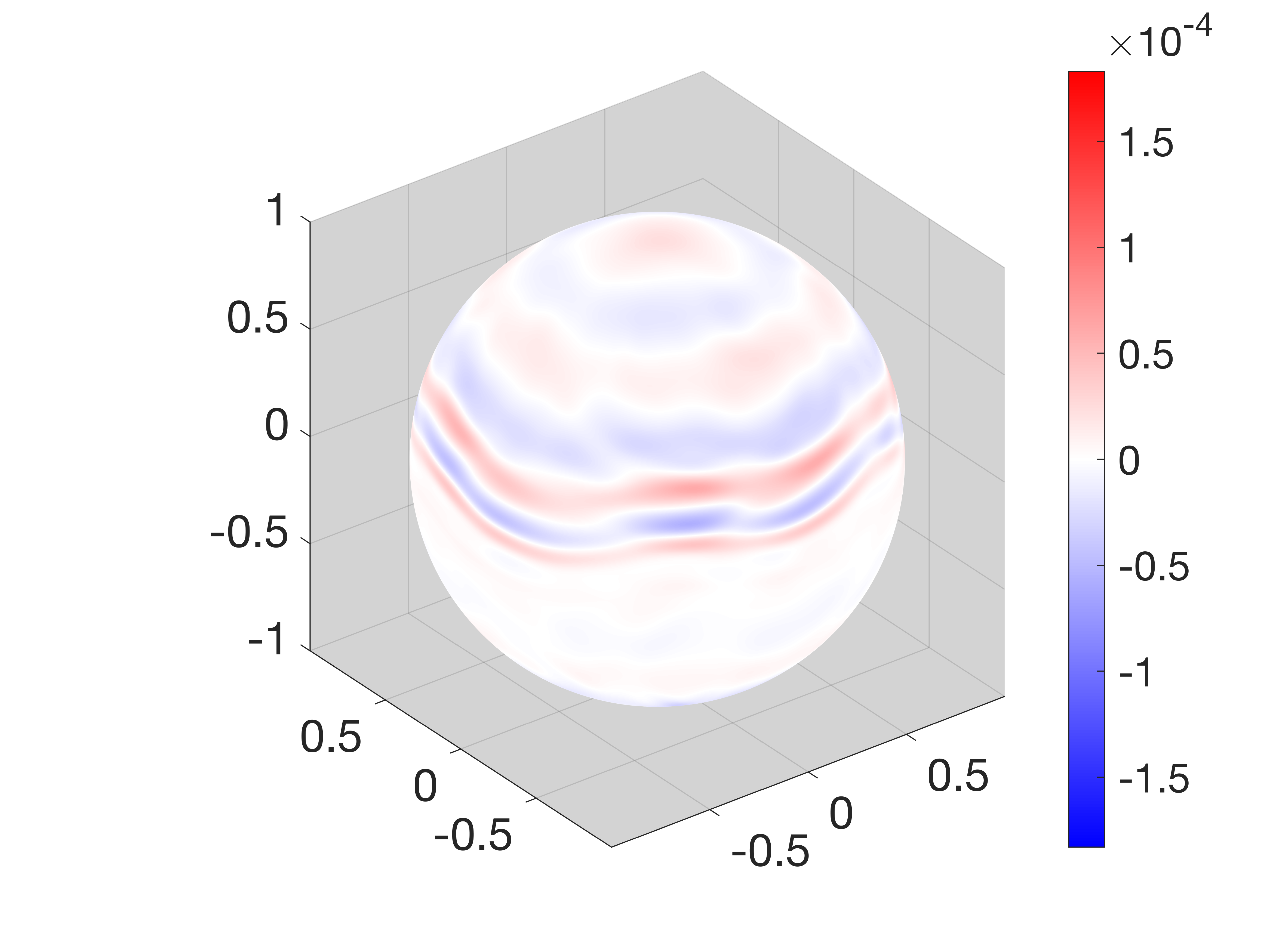}
\caption{{$v_r$ (in  m/s) }}
\lb{fig:SWsimulation_ur}\end{subfigure}
\caption{
Panels~\ref{fig:magnitude}-\ref{fig:SWsimulation_ur} show velocity magnitude, radial vorticity, 3D divergence, and radial velocity, respectively. 
In~\ref{fig:SWsimulation_ur}, $v_r$ is constructed to satisfy $\Divt\bv=0$: $v_r(a_E)=0$ and 
$v_r(a_E+\delta r)\approx-(\grad\bdot\bu)\,\delta r$ on a shell 
$\delta r=10^{-6}a_E$ above the surface. The embedding 3D Euclidean space is shown in units of $a_E$.
}
\label{fig:SWsimulation}
\end{figure}

Fig.~\ref{fig:SWsimulation} shows simulation results. Kinetic energy peaks at approximately day 6 as the instability generates eddying motions. Below, we use the velocity field on day 8, visualized in Fig.~\ref{fig:SWsimulation}. 
The radial velocity in Fig.~\ref{fig:SWsimulation_ur}, constructed as
$v_r \approx -(\grad\bdot\bu)\,\delta r$ (see caption), is a proxy for the lateral divergence and shows the gravity waves; the vorticity in
Fig.~\ref{fig:SWsimulation_vort} highlights the eddies.
The 3D divergence in Fig.~\ref{fig:SWsimulation_div} presents one of the benchmarks, cf.~eq.~\eqref{eq:Div3D}, against which we compare the divergence of the filtered velocity below.

\begin{figure}[h]
\centering
\begin{subfigure}[t]{0.24\textwidth}
\includegraphics[width=\textwidth]{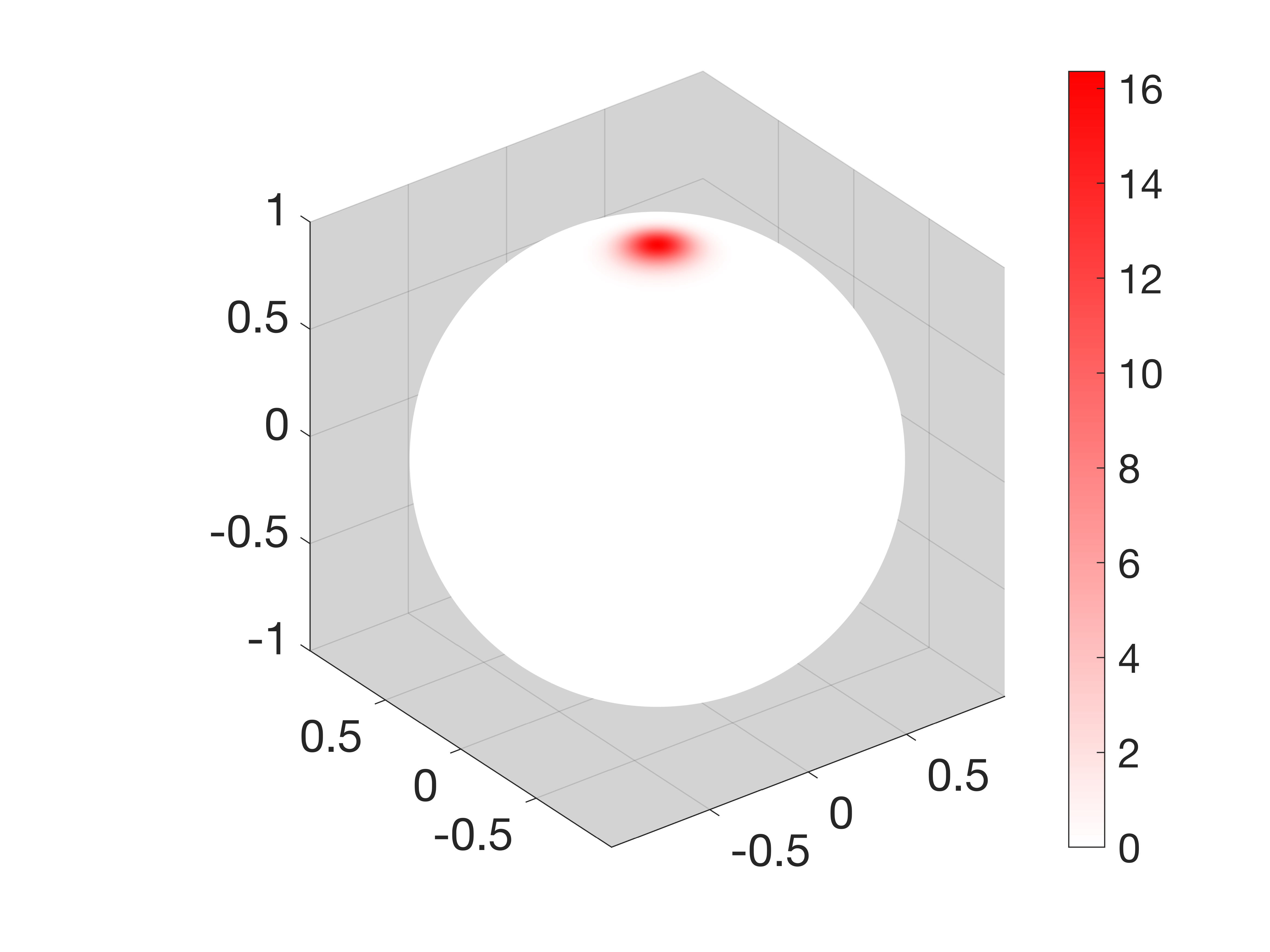}
\caption{Kernel: $a_E^2\,G$}
\label{fig:GaussianKernel}
\end{subfigure}
\begin{subfigure}[t]{0.24\textwidth}
\includegraphics[width=\textwidth]{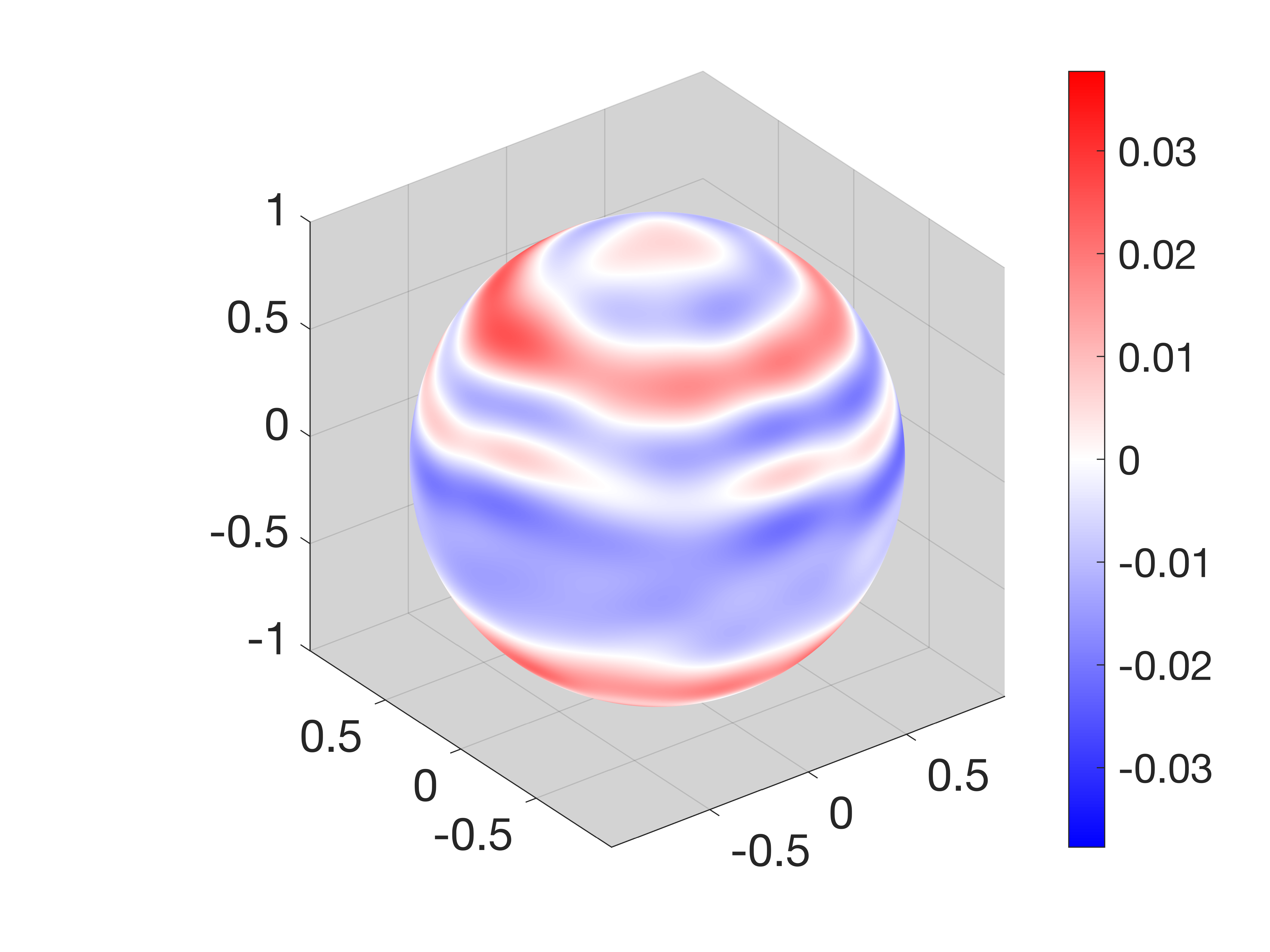}
\caption{$\Divt(G\cConv\bv)$: Ambient}
\label{fig:Div3D_SW_Direct}
\end{subfigure}
\begin{subfigure}[t]{0.24\textwidth}
\includegraphics[width=\textwidth]{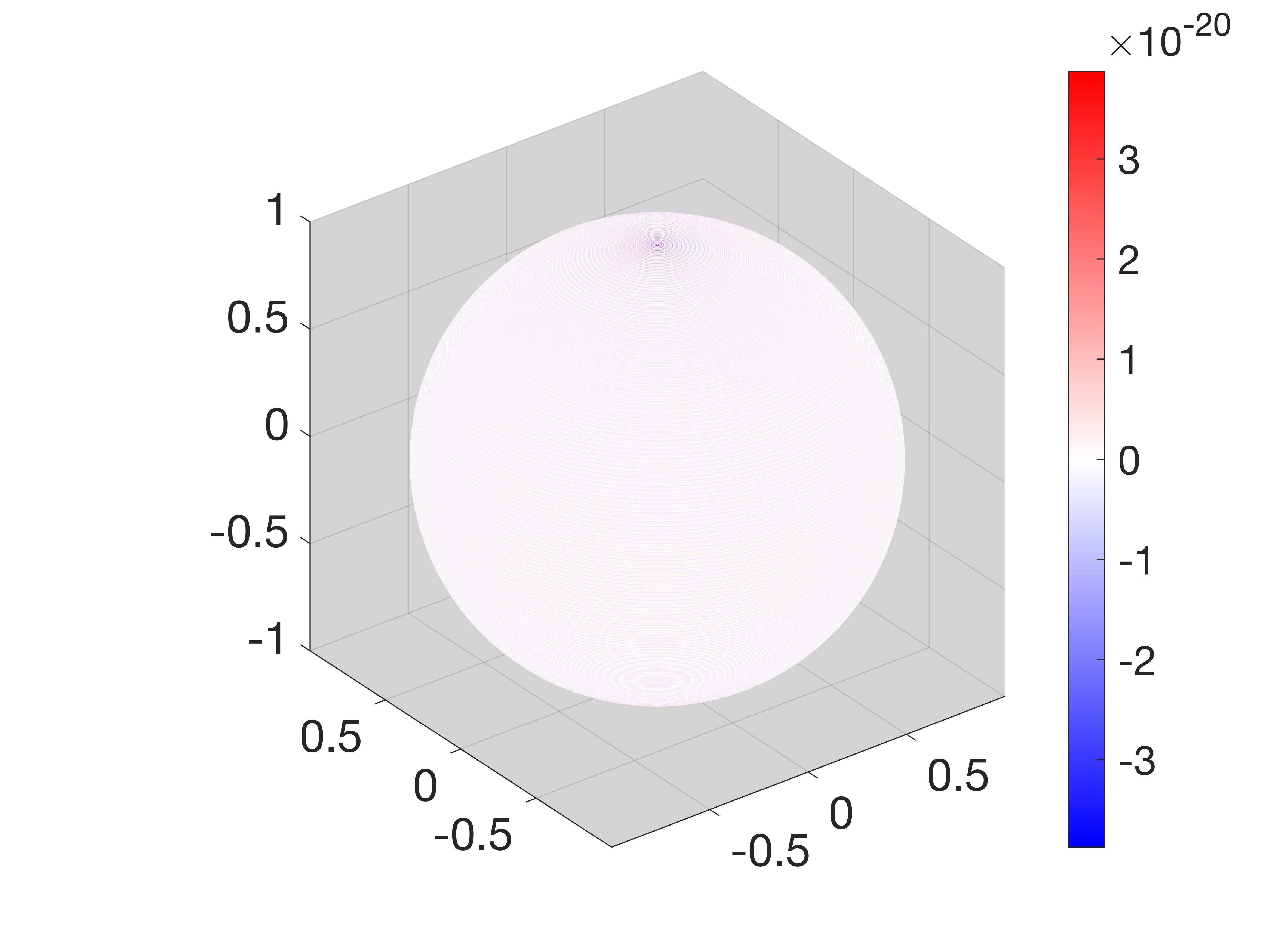}
\caption{$\Divt(G\,\vConv\bv)$: Edmonds}
\label{fig:Div3D_SW_Edmonds}
\end{subfigure}
\begin{subfigure}[t]{0.24\textwidth}
\includegraphics[width=\textwidth]{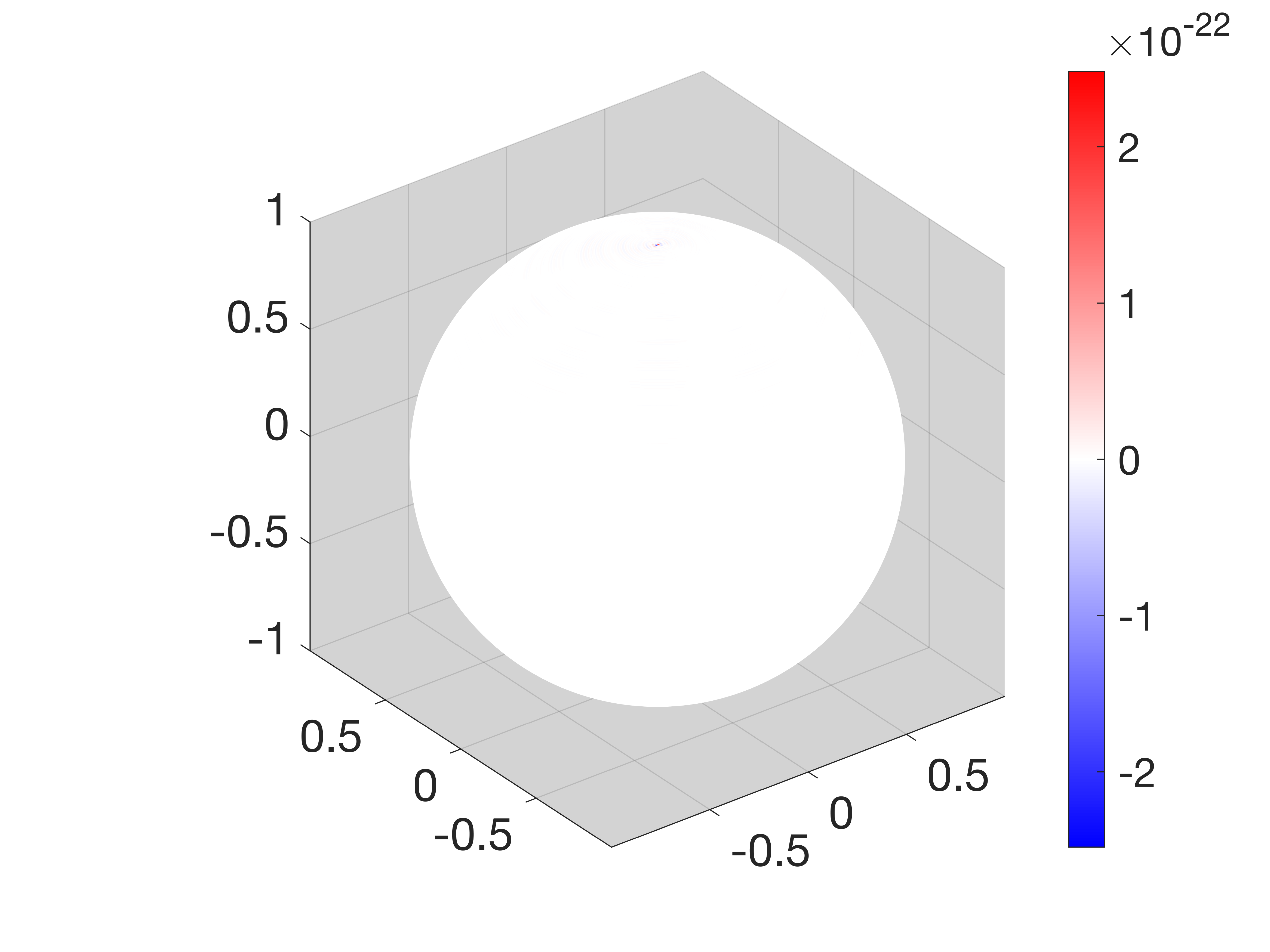}
\caption{$\Divt(G\,\vConv\bv)$: Helmholtz}
\label{fig:Div3D_SW_Helmholtz}
\end{subfigure}
\caption{
Panel~\ref{fig:GaussianKernel}: normalized Gaussian kernel $G$ (width ${\approx}\,1000$~km), shown in dimensionless form $a_E^2\,G$. The generalized convolution $G\vConv\bv$ is evaluated two ways: panel~\ref{fig:Div3D_SW_Edmonds} via the Edmonds basis (eq.~\eqref{eq:GeneralConvDef}), and panel~\ref{fig:Div3D_SW_Helmholtz} via the Helmholtz scalars (eq.~\eqref{eq:GeneralVectorConv}). That both yield $\Divt(G\vConv\bv)=0$ to within round-off is nontrivial and demonstrates the results in \cite{Aluie19}. Panel~\ref{fig:Div3D_SW_Helmholtz} values are even smaller than those in Fig.~\ref{fig:SWsimulation_div} because of filtering. In contrast, the ambient convolution (panel~\ref{fig:Div3D_SW_Direct}) yields a spurious divergence roughly eighteen orders of magnitude larger.
} 
\label{fig:Div3D_SW}
\end{figure}

All convolutions below use the kernel in
Fig.~\ref{fig:GaussianKernel}. Since the flow satisfies eq.~\eqref{eq:Div3D}, a commutative filter must return a filtered velocity that is also 3D divergence-free;
this is a stringent test because the unfiltered $\Divt\bv$ is at double-precision round-off by construction, $O(10^{-21})\,\mathrm{s}^{-1}$ (Fig.~\ref{fig:SWsimulation_div}). Figure~\ref{fig:Div3D_SW}
shows that the generalized convolution passes this test: evaluating
$G\vConv\bv$ either via the Edmonds basis in eq.~\eqref{eq:GeneralConvDef} (Fig.~\ref{fig:Div3D_SW_Edmonds}) or via the Helmholtz scalars in eq.~\eqref{eq:GeneralVectorConv} (Fig.~\ref{fig:Div3D_SW_Helmholtz}) yields a divergence of $O(10^{-22}\!-\!10^{-20})\,\mathrm{s}^{-1}$, i.e., at numerical accuracy.
In contrast, the ambient convolution (Fig.~\ref{fig:Div3D_SW_Direct}) produces $\Divt(G\cConv\bv)=O(10^{-2})~$s$^{-1}$: an error more than eighteen orders of magnitude larger, and, importantly, two to three orders of magnitude larger than the flow's lateral divergence (Fig.~\ref{fig:SWsimulation_ur}) and vorticity (Fig.~\ref{fig:SWsimulation_vort}). The culprit is the ``radial-lateral mixing'' by the ambient convolution, as illustrated in Fig.~\ref{fig:SphericalFiltering_CounterPoloidal} and explained spectrally in eqs.~\eqref{eq:EdmondsEigen}-\eqref{eq:EdmondsTransf}. 
Since $\Divt$ is evaluated with impermeability, $v_r(a_E)=0$, imposed at the surface, the leaked radial velocity, reaching ${\sim}0.2$~m/s, appears as a spurious divergence ${\sim}|\hBe_r\bdot(G\cConv\bv)|/\delta r$.
The componentwise convolution (eq.~\eqref{eq:FuzzyConvVectors}), which filters the spherical components as scalars, cannot leak into the radial direction; it yields
$\Divt$ errors of $O(10^{-5})\,\mathrm{s}^{-1}$ (not shown), which are still significant and of the order of $\grad\bdot\bu$ (Fig.~\ref{fig:SWsimulation_ur}), though far from the ambient convolution's catastrophic failure.

\begin{figure}[h]
\centering
\begin{subfigure}[t]{0.24\textwidth}
\includegraphics[width=\textwidth]{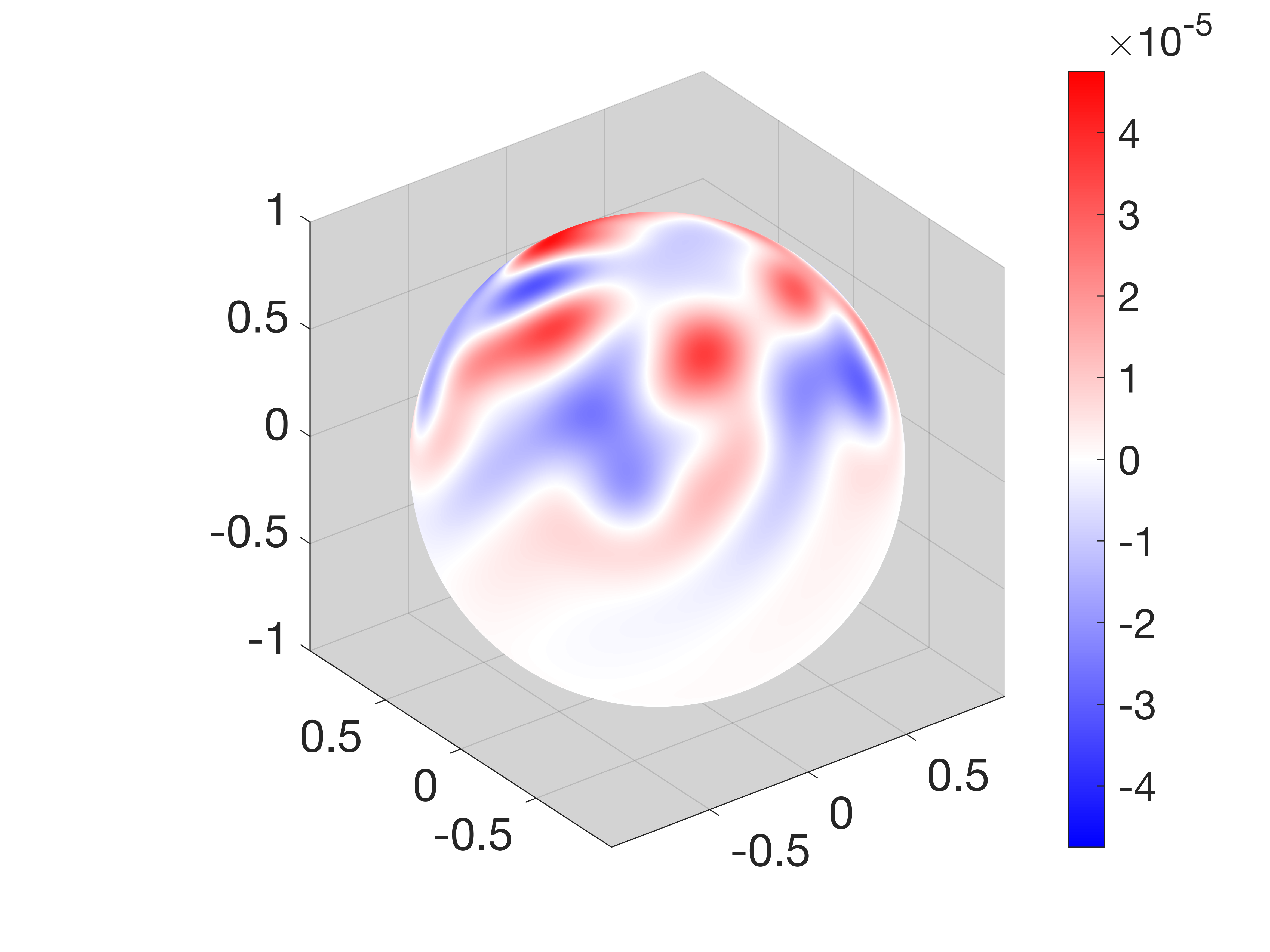}
\caption{{$G\cConv \omega_r$}}
\label{fig:FilteredVorticity}
\end{subfigure}
\begin{subfigure}[t]{0.24\textwidth}
\includegraphics[width=\textwidth]{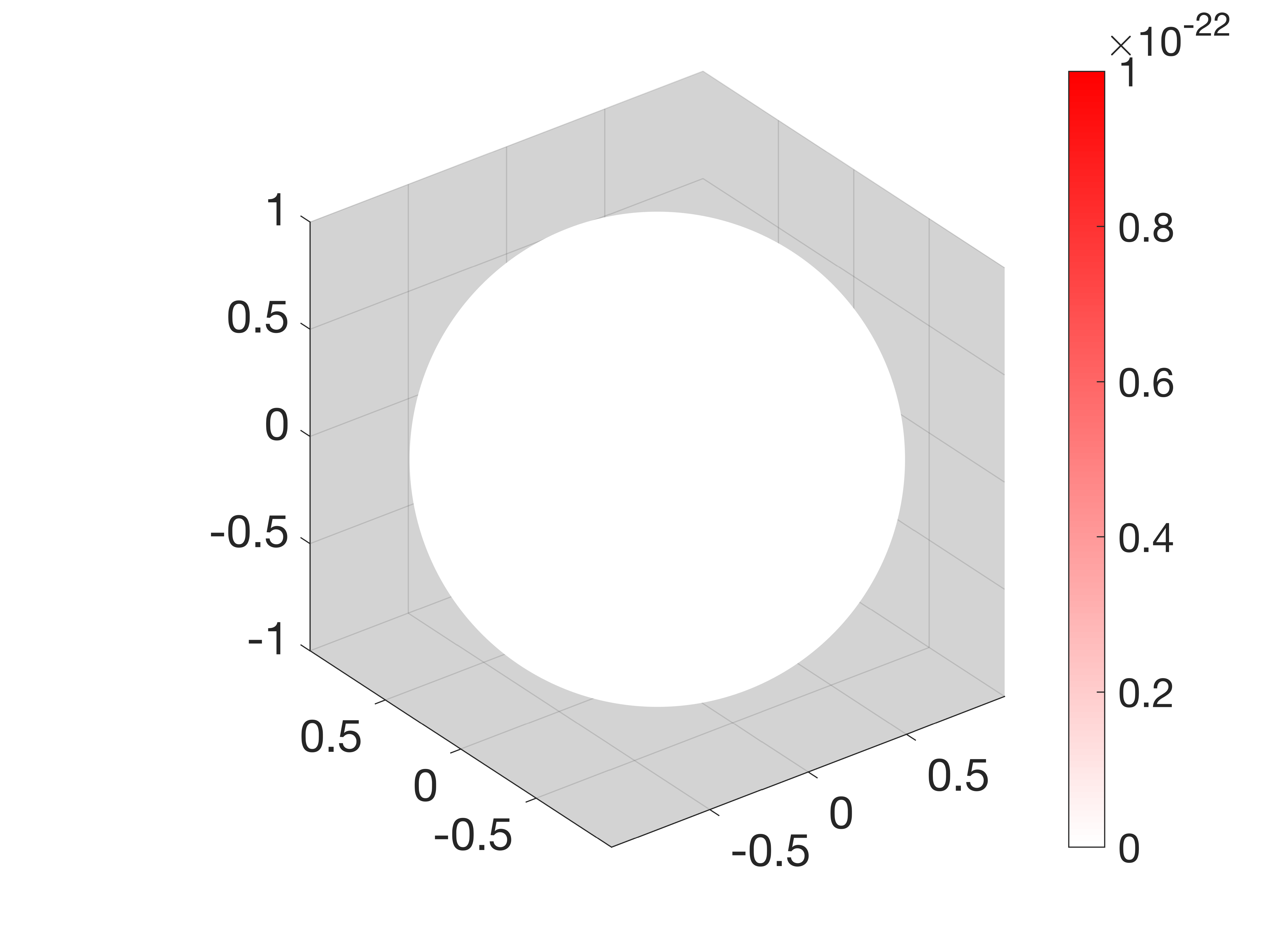}
\caption{{$|(\grad\btimes\textrm{Helm})\bdot\hBe_r - G\cConv \omega_r|$}}
\label{fig:Vort_SW_Helmholtz}
\end{subfigure}
\begin{subfigure}[t]{0.24\textwidth}
\includegraphics[width=\textwidth]{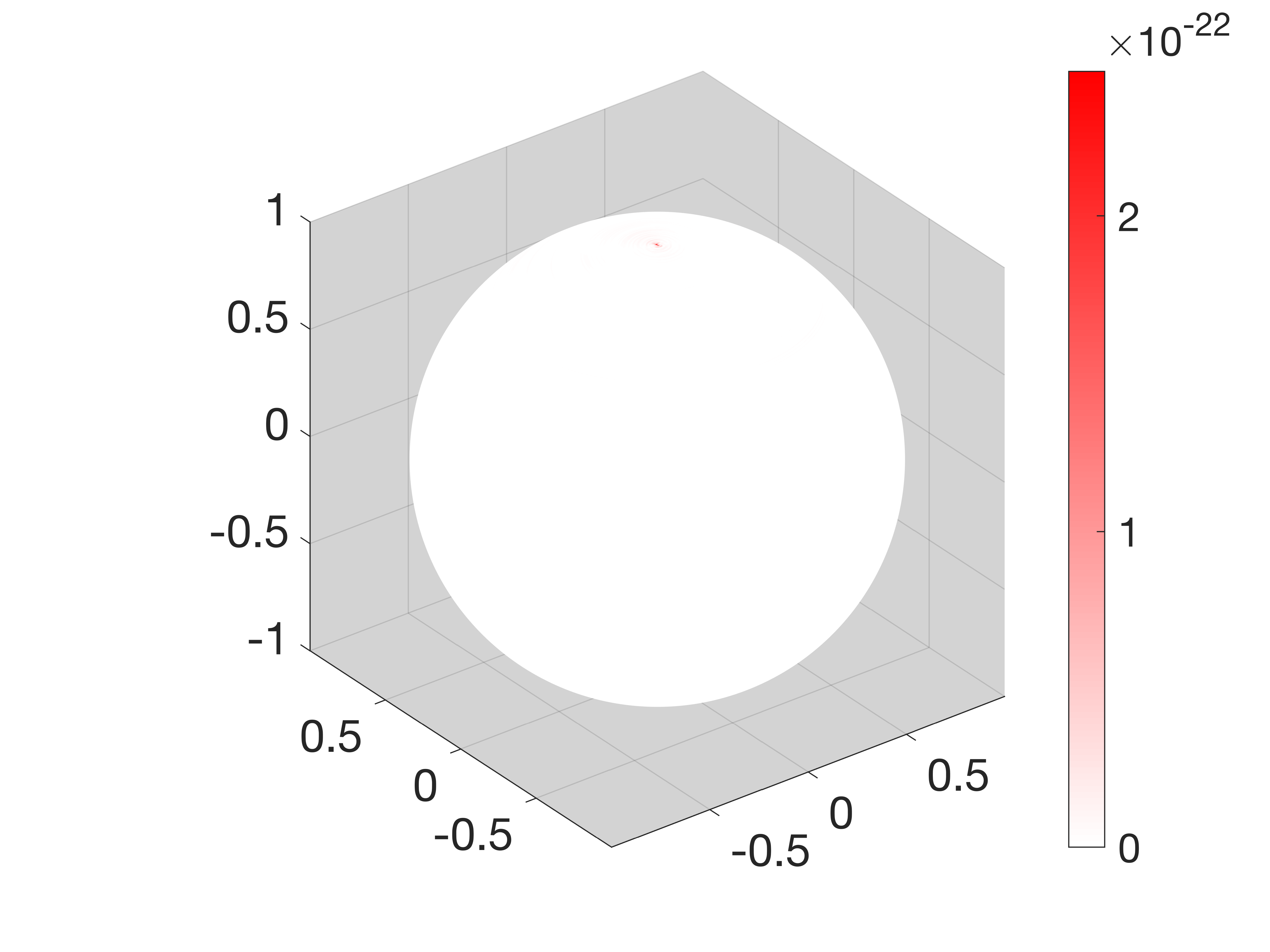}
\caption{{$|(\grad\btimes\textrm{Edmo})\bdot\hBe_r - G\cConv \omega_r|$}}
\label{fig:Vort_SW_Edmonds}
\end{subfigure}
\begin{subfigure}[t]{0.24\textwidth}
\includegraphics[width=\textwidth]{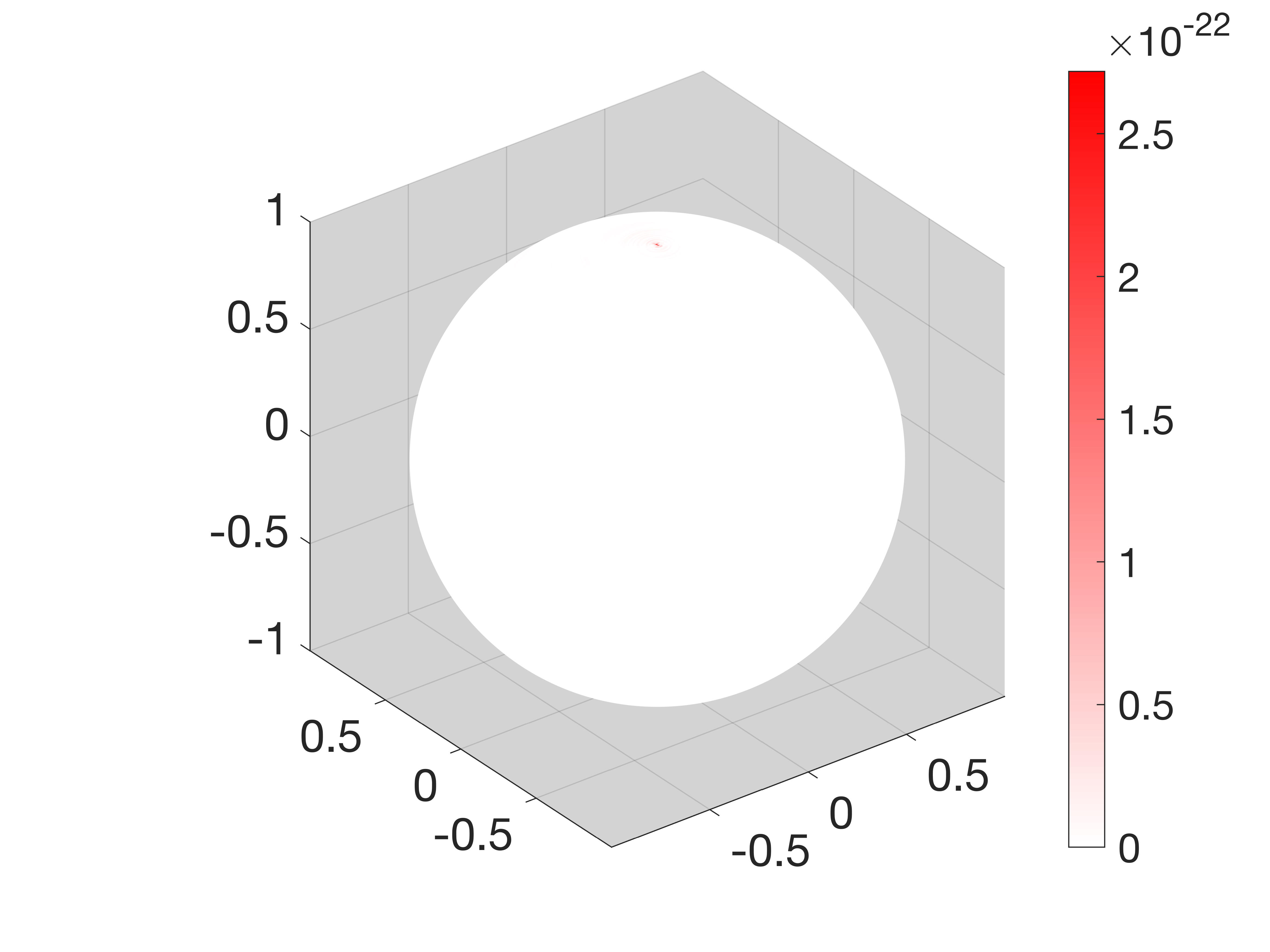}
\caption{{$|(\grad\btimes\textrm{Amb})\bdot\hBe_r - G\cConv \omega_r|$}}
\label{fig:Vort_SW_Ambient}
\end{subfigure}
\caption{Ambient convolution is identical to the generalized convolution when applied to a laterally solenoidal vector field, which accounts for the flow's radial vorticity. Panel~(\ref{fig:FilteredVorticity}) shows the spherical convolution of the radial vorticity as a scalar, the benchmark. Panels~(\ref{fig:Vort_SW_Helmholtz}),(\ref{fig:Vort_SW_Edmonds}),(\ref{fig:Vort_SW_Ambient}) calculate the radial vorticity from the filtered vector field, with the vector filtering done via the (\ref{fig:Vort_SW_Helmholtz}) Helmholtz scalars, the (\ref{fig:Vort_SW_Edmonds}) Edmonds transformation, and the (\ref{fig:Vort_SW_Ambient}) ambient convolution.
Ambient convolution agrees with~(\ref{fig:FilteredVorticity}) to round-off because the solenoidal basis functions are eigenfunctions of the ambient convolution (eq.~\eqref{eq:NOTeigenbasis_3}).
}
\label{fig:SW_FilterVort}
\end{figure}

Figure~\ref{fig:SW_FilterVort} probes the (laterally) solenoidal flow via the radial vorticity. 
Since $\bPhi_{n,j}$ is an eigenfunction of both $G\cConv$ (eq.~\eqref{eq:NOTeigenbasis_3}) and $G\vConv$, both convolutions commute with curl: vorticity of filtered velocity equals $G\cConv\omega_r$ to round-off.

\begin{figure}[h]
\centering
\begin{subfigure}[t]{0.24\textwidth}
\includegraphics[width=\textwidth]{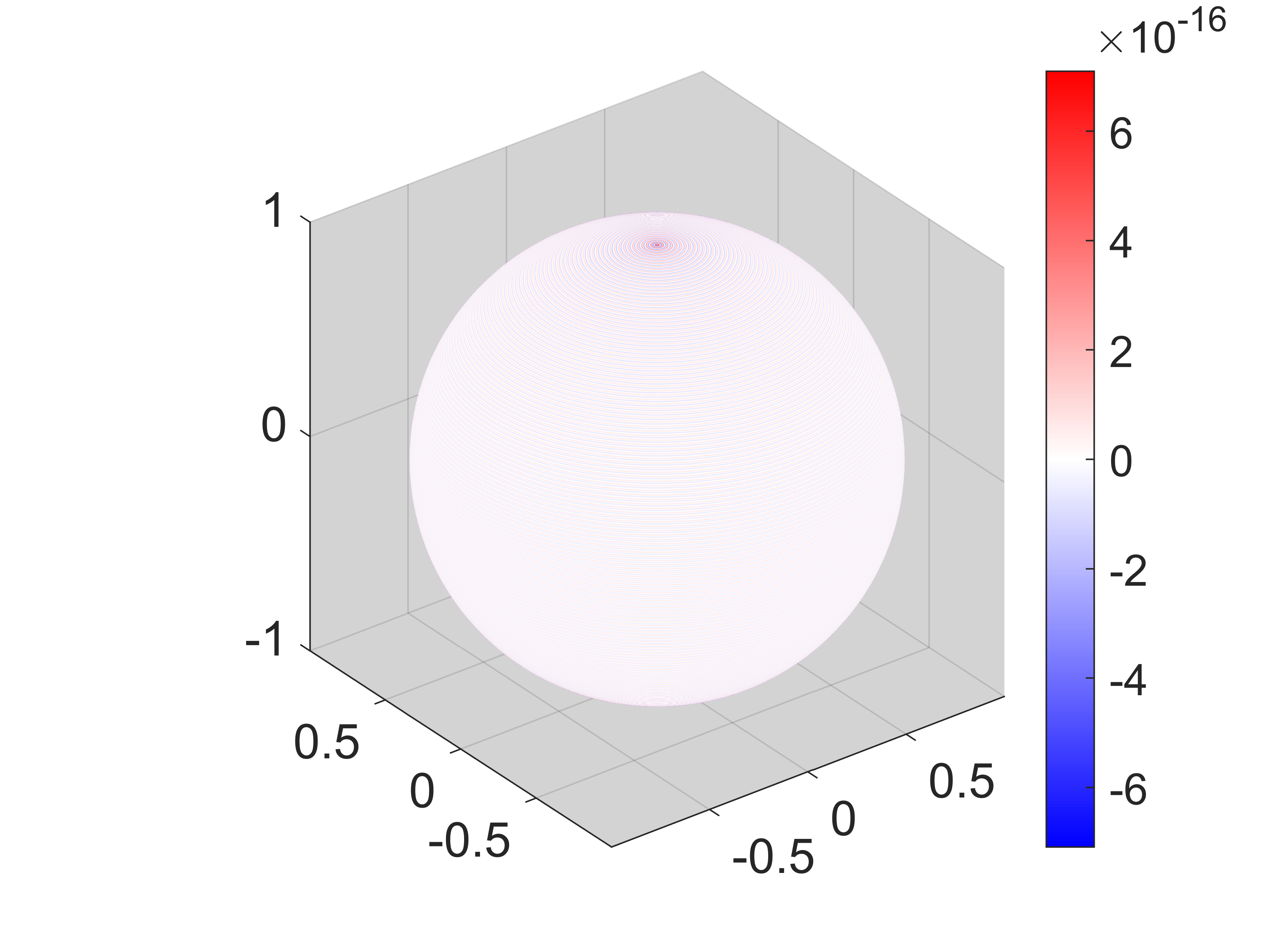}
\caption{{Edmonds: $[G\vConv\grad h]\bdot\hBe_r$}}
\label{fig:GradientEdmondsRadial}\end{subfigure}
\begin{subfigure}[t]{0.24\textwidth}
\includegraphics[width=\textwidth]{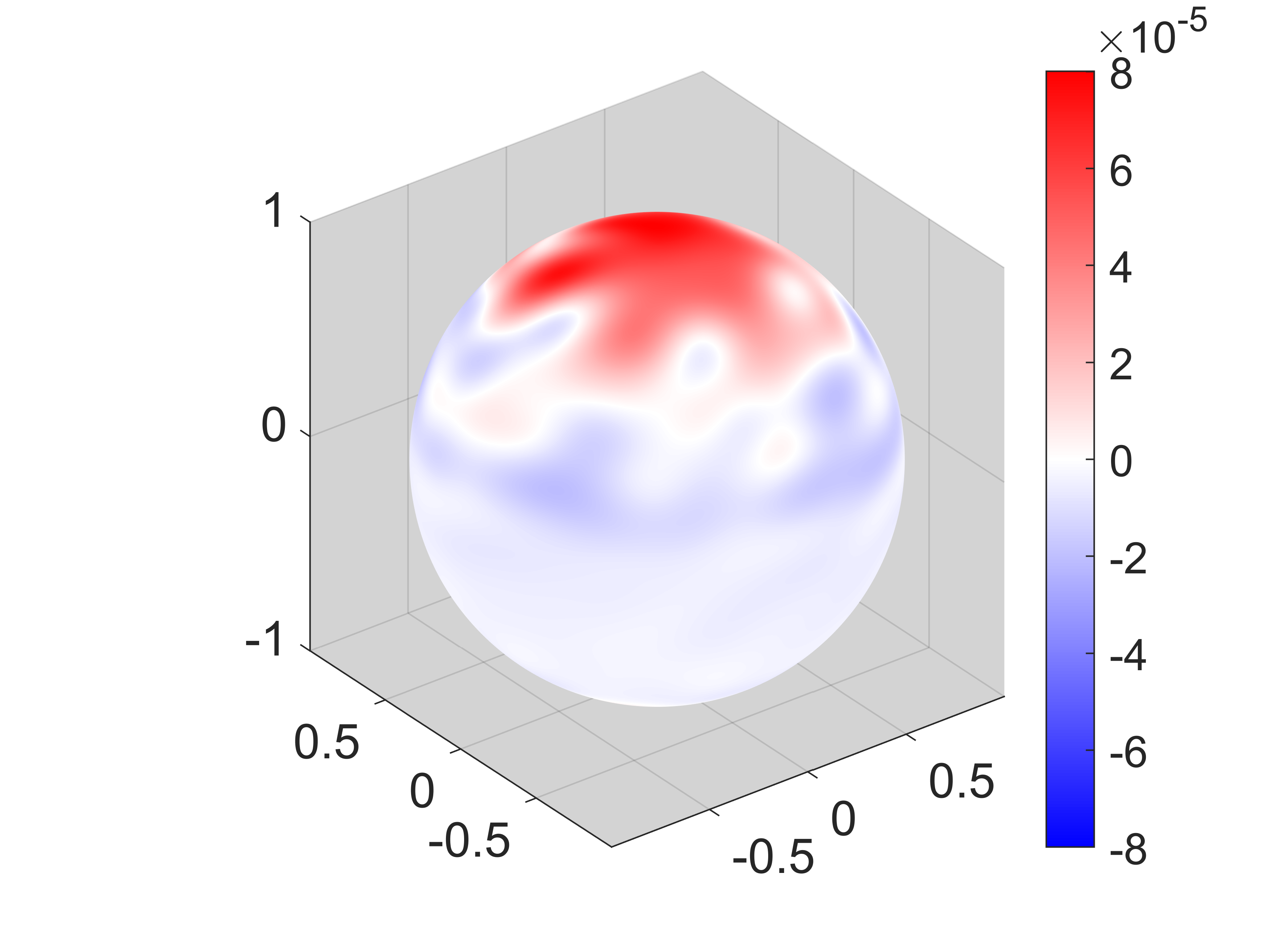}
\caption{{Ambient: $[G\cConv\grad h]\bdot\hBe_r$}}
\label{fig:GradientAmbientRadial}\end{subfigure}
\begin{subfigure}[t]{0.24\textwidth}
\includegraphics[width=\textwidth]{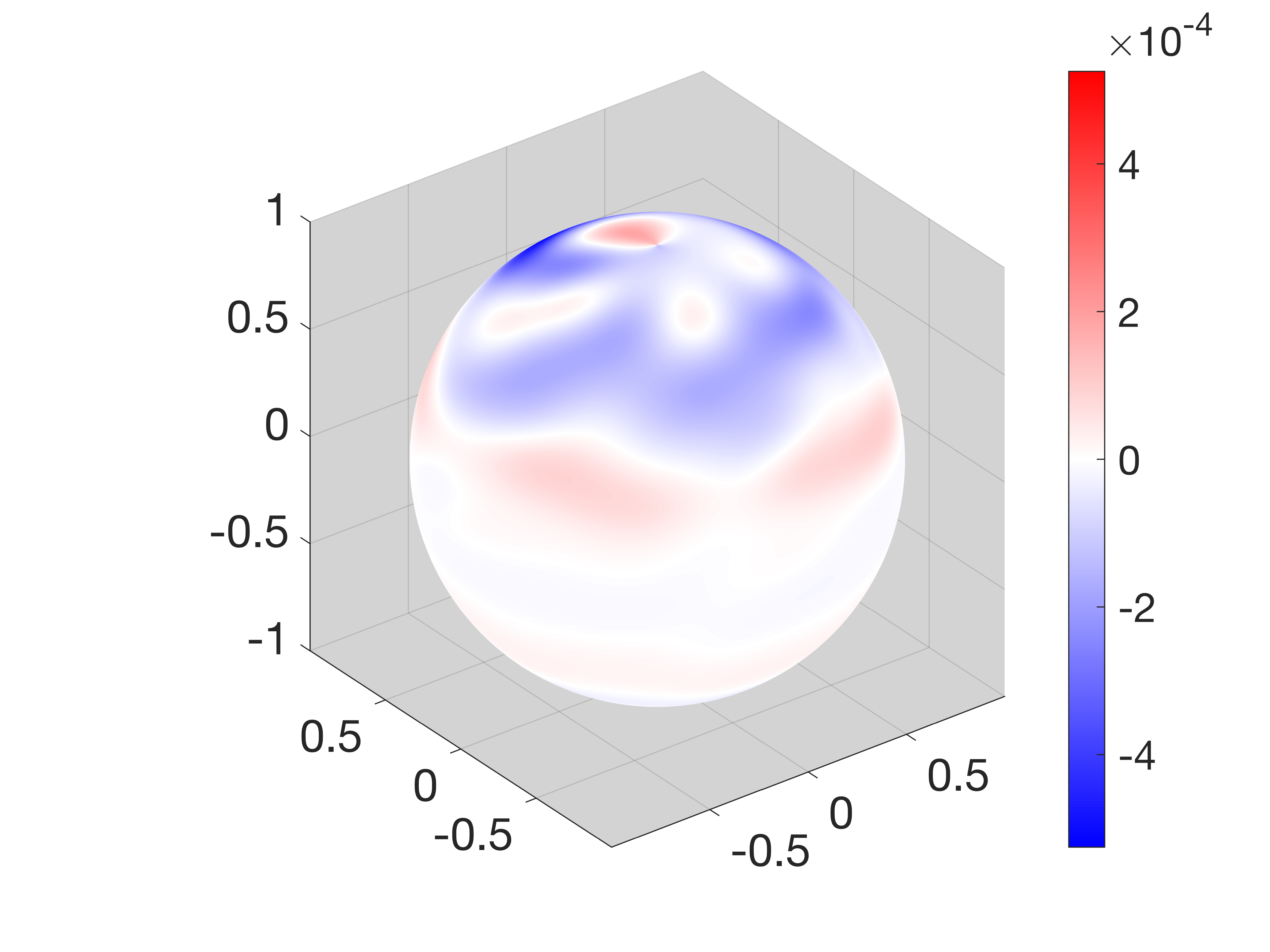}
\caption{{Helmholtz:\,$\hBe_\phi\bdot\grad(G\cConv h)$}}
\label{fig:GradientHelmholtzNorth}\end{subfigure}
\begin{subfigure}[t]{0.24\textwidth}
\includegraphics[width=\textwidth]{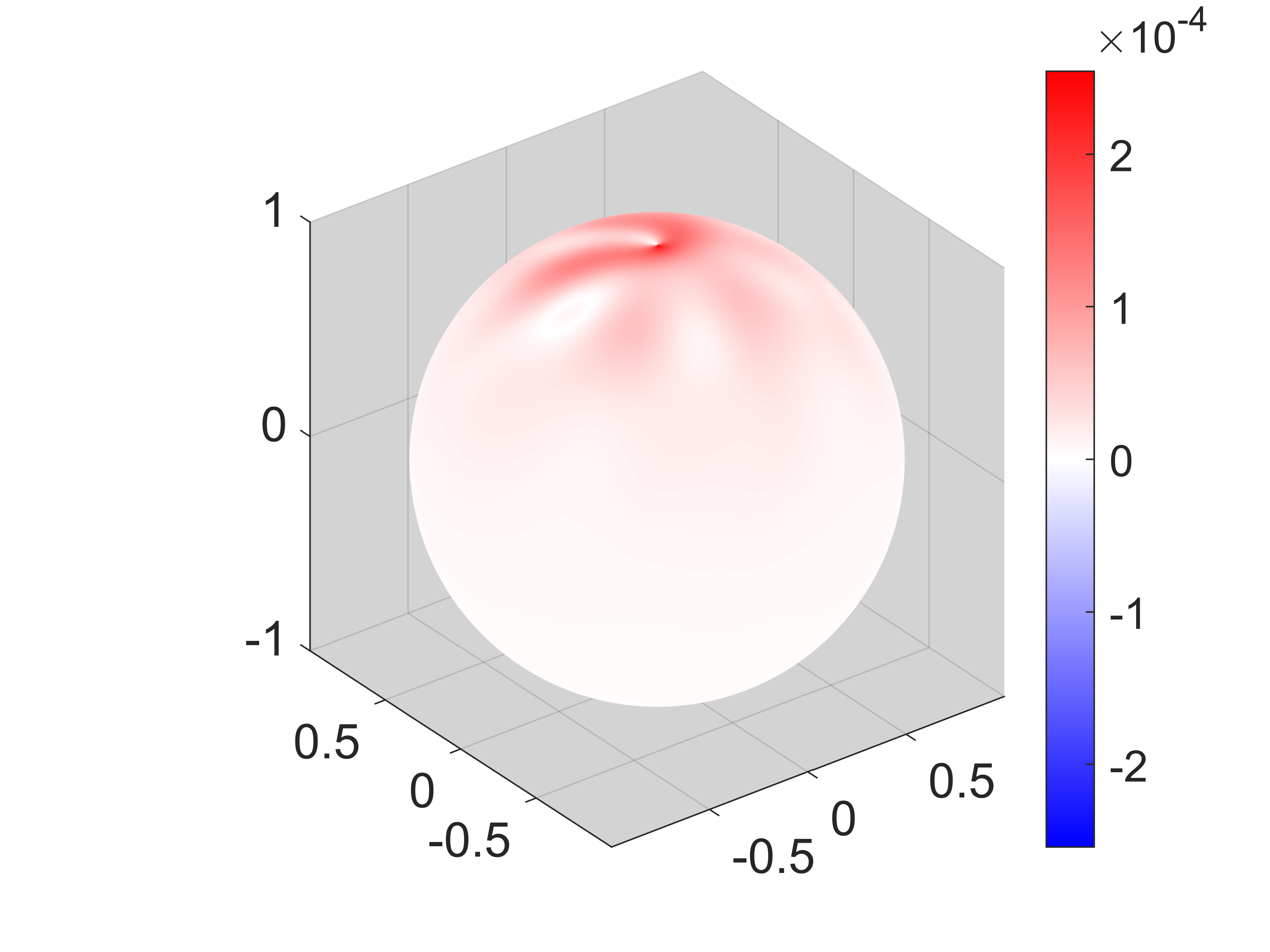}
\caption{Helmholtz -- Cmptwise}
\label{fig:GradientHelmholtz-Cmptwise-north}\end{subfigure}
\caption{Panel~(\ref{fig:GradientEdmondsRadial}):
the tangent gradient vector,  filtered using the generalized convolution by an Edmonds transformation, remains tangent to within round-off. In contrast, 
(\ref{fig:GradientAmbientRadial}) shows that the ambient convolution yields a spurious radial component eleven orders of magnitude larger than~(\ref{fig:GradientEdmondsRadial}), reaching ${\sim}20\%$ of the gradient magnitude in~(\ref{fig:GradientHelmholtzNorth}).
Panel~(\ref{fig:GradientHelmholtzNorth}): meridional component of the generalized convolution of the height gradient, evaluated via the Helmholtz scalars.
Panel~(\ref{fig:GradientHelmholtz-Cmptwise-north}): commutation error, $\big|\{\grad(G\cConv h)\}\bdot\hBe_\phi - [G\fuzConv\grad h]\bdot\hBe_\phi\big|$, of the componentwise convolution in the meridional gradient, comparable to the magnitude of the gradient itself in~(\ref{fig:GradientHelmholtzNorth}).
} 
\label{fig:SW_FilterGrad}
\end{figure}

Figure~\ref{fig:SW_FilterGrad} tests commutation with first-order derivatives using the height gradient $\grad h$, a tangent vector field whose Helmholtz potential is $h$ itself, so that the generalized
convolution satisfies $G\vConv(\grad h)=\grad(G\cConv h)$  (eq.~\eqref{eq:GeneralVectorConv}). Figs.~\ref{fig:GradientEdmondsRadial},\ref{fig:GradientAmbientRadial}
probe whether filtering preserves tangency. The generalized convolution (Fig.~\ref{fig:GradientEdmondsRadial}) yields a zero radial component to round-off, with relative error of $O(10^{-12})$ compared to the gradient magnitude (Fig.~\ref{fig:GradientHelmholtzNorth}). In comparison, the ambient convolution (Fig.~\ref{fig:GradientAmbientRadial}) yields a radial component eleven
orders of magnitude larger, reaching $\sim\!20\%$ of the gradient magnitude, which was expected from eq.~\eqref{eq:RadialTangentFail} and illustrated in Fig.~\ref{fig:SphericalFiltering_CounterPoloidal}. Figure~\ref{fig:GradientHelmholtz-Cmptwise-north} shows the commutation error of the componentwise convolution in the meridional
gradient: the error is $O(1)$ relative to the true filtered gradient in Fig.~\ref{fig:GradientHelmholtzNorth}, i.e., comparable to the signal itself. Note that divergences and gradients
such as $\grad h$ and $\grad\bdot (h\bu)$ appear in the governing equations~\eqref{eq:SWequations}, so commutation errors of this kind directly contaminate any coarse-grained dynamics built on non-commutative filters. 

\acknowledgments
HA thanks B.~Storer, S.~Kouhen, and P.~H.~Sichani for helpful discussions. HA is supported by NSF grants OCE-2123496, OCE-2446475, TI-2449340, PHY-2020249, PHY-2206380, OIA-2535317, DOE grant DE-SC0020229, and NNSA grants DE-NA0004144, 
 DE-NA0004134. Computing time was provided by NERSC and TACC.
\bibliographystyle{eplbib}
\bibliography{Refs}

\newpage
\clearpage
\section{Supplementary Material (SM)}
\setcounter{figure}{0}  
\renewcommand{\thefigure}{S\arabic{figure}}
\renewcommand{\thetable}{S\arabic{table}}
\,\vspace{\bigskipamount}

\textbf{Uniqueness of the commutative convolution.}
The generalized convolution~\eqref{eq:GeneralConvDef} is the \emph{only} linear operator compatible with a given scalar kernel  under the commutation requirements. Precisely, let $G$ be an integrable scalar function on $\cS$ that is  axisymmetric, let $G\cConv$ be the spherical convolution of scalars in eq.~\eqref{eq:ConvolutionScalar} (or eq.~\eqref{eq:ScalarEigen}), and let $G_1$ be any linear operator on sufficiently smooth vector fields $\bv = v_r\hBe_r + \bu$ on $\cS$ (with $\bu$
tangential) satisfying
\begin{subequations}\lb{eq:UniqueConds}
\begin{align}
\hBe_r\bdot(G_1\bv) &= G\cConv v_r,\lb{eq:UniqueConds-1}\\ 
\grad\bdot(G_1\bv)_t &= G\cConv(\grad\bdot\bu),\lb{eq:UniqueConds-2}\\
\hBe_r\bdot[\grad\btimes(G_1\bv)_t] &= G\cConv\omega_r, \lb{eq:UniqueConds-3}
\end{align}
\end{subequations}
where $(\cdot)_t$ denotes the tangential part. Then $G_1 = G\vConv$ on smooth fields.

The proof is as follows. Decompose $\bu = \grad\Phi + \hBe_r\btimes\grad\Psi$ into Helmholtz scalars, which exist and are unique up to constants on $\cS$. Write
$G_1\bv = v_r'\,\hBe_r + \grad\Phi' + \hBe_r\btimes\grad\Psi'$. Relation~\eqref{eq:UniqueConds-1} gives $v_r' = G\cConv v_r$. Since
$\grad\bdot\bu = \Delta\Phi$ and the scalar convolution commutes with the
scalar Laplacian (both are diagonal in $Y_{n,j}$ by eq.~\eqref{eq:ScalarEigen}), relation~\eqref{eq:UniqueConds-2} yields $\Delta\Phi' = G\cConv\Delta\Phi = \Delta(G\cConv\Phi)$, so
$\Phi' - G\cConv\Phi$ is harmonic on $\cS$, hence constant, and $\grad\Phi' = \grad(G\cConv\Phi)$. 
An identical argument applies to relation~\eqref{eq:UniqueConds-3}: using $\omega_r = \Delta\Psi$ gives
$\hBe_r\btimes\grad\Psi' = \hBe_r\btimes\grad(G\cConv\Psi)$. Therefore $G_1\bv$ coincides with the right-hand side of
eq.~\eqref{eq:GeneralVectorConv} for every $\bv$, i.e.\ $G_1 = G\vConv$.
Conversely, $G\vConv$ satisfies relations~\eqref{eq:UniqueConds} by design \cite{Aluie19S}.

Remark 1: The argument does not assume any specific structure (or kernel representation) of $G$, beyond axisymmetry and integrability\footnote{Even integrability can be relaxed. See Remark~3.}. 
For example, it does not require spatial locality, or positivity, or low-pass character, or preservation of constants. Any linear operator on (sufficiently smooth) vector fields satisfying eq.~\eqref{eq:UniqueConds} is necessarily $G\vConv$. Examples include spectral (triangular) truncation, isotropic wavelet filters, diffusion, smoothed derivatives (band-pass kernels), Green's-function inversion, and filters learned from data.

Remark 2: Smoothness of $\bv$ is assumed only so that every quantity in relations~\eqref{eq:UniqueConds} is classically defined. We can give up smoothness of $\bv$ if we instead require $G_1$ to be a \emph{continuous} linear operator. The argument is as follows. We have for any integrable kernel, $|\dhat{G}(n)| \le \|G\|_{L^1}$ (for a normalized non-negative kernel, we even have $\sup_n|\dhat{G}(n)| = 1$), so $G\vConv$ is a bounded operator on vector fields $\bv\in\mathbf{L}^2(\cS)$ (finite-energy fields, possibly rough). Boundedness of $G\vConv$ implies that it is a continuous operator on $\mathbf{L}^2(\cS)$ \cite{lieb2001analysisS}. Since (i) $G_1$ and $G\vConv$ agree on smooth fields by the result above, (ii) smooth fields are dense in $\mathbf{L}^2(\cS)$ \cite{lieb2001analysisS}, and (iii) both operators are continuous, we deduce that $G_1 = G\vConv$ on all of $\mathbf{L}^2(\cS)$. This broader setting is relevant for turbulent flows in the infinite Reynolds number limit, whose velocity fields need not be smooth. There, coarse-graining serves as a regularization \cite{eyink2008dissipativeS}.

Remark 3: Integrability of $G$ is not needed for the results on smooth ($C^\infty$) vector fields. The argument uses only that $G\circledast$, as defined by eq.~\eqref{eq:ScalarEigen}, is diagonal in $Y_{n,j}$ with eigenvalues $\dhat{G}(n)$ depending only on the degree $n$ (and independent of $j$). This holds for any axisymmetric distribution $G$ acting on smooth fields (equivalently, any spectral multiplier $\dhat G(n)$ that does not grow faster than algebraically with $n\to\infty$). This includes $G=\delta$ (which has $\dhat G(n)=1$ for all $n$), high-pass kernels $\delta - G_{\mathrm{lowpass}}$, and differential operators such as $\Delta$. In this setting, Remark~2 requires only $\sup_n|\dhat{G}(n)|<\infty$, which excludes $\Delta$ but still allows for $\delta$.

Remark 4: Replacing conditions~\eqref{eq:UniqueConds-2},\eqref{eq:UniqueConds-3} with the condition $[\Delta^{\rm vec}, G_1]=0$ is not sufficient to imply $G_1 = G\vConv$, since $\Delta^{\rm vec}$ has the degenerate eigenvalue $-n(n+1)/a^2$ across the radial, potential, and solenoidal components and cannot distinguish them. Even the stronger assumption that the operator $G_1$ is rotation-equivariant (in addition to condition~\eqref{eq:UniqueConds-1}) is not sufficient to show $G_1 = G\vConv$. A counterexample is $G_1: \bv \mapsto (G\cConv v_r)\,\hBe_r + \hBe_r\btimes\bu$, which is rotation-equivariant and satisfies condition~\eqref{eq:UniqueConds-1}, yet violates relations~\eqref{eq:UniqueConds-2},\eqref{eq:UniqueConds-3}; since $G\vConv$ satisfies all of relations~\eqref{eq:UniqueConds} by design, $G_1 \neq G\vConv$.

Remark 5: Forthcoming work extends commutative convolutions to arbitrary manifolds \cite{storer2026inprepS}.

\providecommand{\refname}{References}
\renewcommand{\refname}{\textbf{Supplementary references}}

\end{document}